\documentclass[twocolumn,amsmath,amssymb,pra]{revtex4-2}

\usepackage{amssymb,amsfonts,amsmath}
\usepackage{graphicx,epsfig,psfrag}
\usepackage{color}
\usepackage{natbib}
\usepackage{url}
\usepackage[breaklinks=true]{hyperref}
\usepackage{mathtools}
\usepackage{subfigure}
\hypersetup{
        colorlinks=true,
        citecolor    = blue
}
\usepackage{bookmark}
\usepackage{epic}
\usepackage{longtable}

\usepackage{bbm}
\usepackage{ulem}
\usepackage{braket}
\usepackage{xcolor}
\usepackage{csquotes}
\usepackage{graphicx}
\usepackage{tikz}
\usepackage{extarrows}
\usetikzlibrary{trees}
\usetikzlibrary{decorations.pathmorphing}
\usetikzlibrary{decorations.markings}
\usetikzlibrary{positioning,arrows}
\usetikzlibrary{shapes.arrows}
\usepackage{leftidx}
\usepackage{framed}
\usepackage{tcolorbox}
\usepackage{bm}

\usepackage{float}

\tikzset{->-/.style={decoration={
  markings,
  mark=at position #1 with {\arrow{>}}},postaction={decorate}}}

\newcommand\identity{1\kern-0.25em\text{l}}

\begin{document} 

\title{The impact of noise on the simulation of NMR spectroscopy on NISQ devices}

\author{Andisheh Khedri}
\author{Pascal Stadler}
\author{Kirsten Bark}
\author{Matteo Lodi}
\author{Rolando Reiner}
\author{Nicolas Vogt}
\author{Michael Marthaler}
\author{Juha Lepp\"akangas}
\affiliation{HQS Quantum Simulations GmbH, Rintheimer Str. 23, 76131 Karlsruhe, Germany}


\begin{abstract}

With the surge of quantum computing platforms that continue to push the boundaries of capabilities of noisy intermediate-scale quantum computers, there is a growing interest in finding relevant applications and quantifying the corresponding error budgets.
We present a simulation of nuclear magnetic resonance (NMR) spectroscopy of small organic molecules on publicly available cloud quantum computers. We are using two quantum computing platforms, namely IBM's quantum processors based on superconducting qubits and IonQ's Aria trapped ion quantum computer addressed via Amazon Braket. We analyze the impact of noise on the obtained NMR spectra, and we formulate an effective decoherence rate that quantifies the threshold noise that our proposed algorithm can tolerate. We show that the effective decoherence rate can be calculated using simple fidelity metrics that are available by cloud quantum computing providers. Our investigation paves the way to better employ such application-driven quantum tasks on current noisy quantum devices.

\end{abstract}

\maketitle

\section{Introduction}
Among potential applications for quantum computers,
it is clear that quantum simulation, the goal of solving intrinsically quantum mechanical problems, is one of the most interesting and promising ones.
Digital quantum simulation ~\cite{Fauseweh2024} is often considered
to address open questions in the field of quantum chemistry~\cite{OMalley2016,Kandala2017} and condensed matter physics~\cite{Monroe2021,Zhang2017}.
In this respect, the variational quantum algorithms have been introduced~\cite{Cerezo2021}, as a mean to find the low-temperature properties, i.e., ground and low lying states, but suffer from practical obstacles such as the “barren plateau" ~\cite{Cerezo2021_barren}.

In recent years there has been a growing interest in non-variational approaches to digital quantum simulation of quantum matter,
including the Trotterized time evolution~\cite{Fauseweh2021,Salathe2015,Barends2015}.
The advantage of such approaches over variational algorithms is the proven exponential quantum advantage when simulating a time evolution.
An important topic that requires a quantum mechanical time evolutions is simulation of the nuclear magnetic resonance (NMR) spectroscopy~\cite{Burov2024,Sels2020},
that has been recently studied using trapped-ion quantum computers~\cite{Seetharam2023}.
The NMR problem addresses the time evolution of a many-body system that is inherently computationally hard \cite{Kuprov2014,Kuprov2011}, due to the exponential growth of the computational space with the number of particles.
While powerful classical methods exist that can approximate the NMR spectrum, they can reach their limits for approx. 50 spins and low magnetic fields.
Therefore, advances in the digital quantum simulation with quantum processors would be a valuable breakthrough.

Given that simulating NMR spectrum seems to be an ideal problem for quantum computers, it is of course also an interesting question to ask if
it is feasible to be tackled using the current generation of noisy intermediate scale quantum (NISQ) computers.
While it has been established that a fault-tolerant quantum computer has the potential to speed-up various computational problems~\cite{Montanaro2016} such as factoring large numbers~\cite{Shor1994,Shor1997},
finding applications for which NISQ devices can outperform classical computers has proven to be quite difficult.
The fragility of quantum coherence, owing to various sources of noise, is the main challenge ahead in unlocking the full potential of NISQ computers.
To tackle such a challenge, various platforms have been proposed such as superconducting qubits~\cite{Kjaergaard2020,Krantz2019,Wendin2017}, silicon spin qubits~\cite{Noiri2022}, photonics~\cite{Flamini_2019}, neutral atoms~\cite{Henriet2020}, and trapped ions~\cite{Lanyon2011,Bruzewicz2019}.
Despite all the great advances, near-term quantum computers are prone to decoherence and other noisy events like
crosstalk~\cite{Gambetta2012} and coherent overroation~\cite{Kueng2016}. Scaling these systems to a regime beyond the capabilities of classical computers remains an open challenge~\cite{Preskill2018}.
To this end, various mitigation techniques~\cite{Maciejewski2020,Temme2017}
have been introduced to increase the applicability of current Noisy Intermediate-Scale Quantum (NISQ) devices.
However, the computational overhead of the implementation of these mitigation techniques is a crucial drawback.

The impact of noise on the practicality of simulating a Trotterized time evolution is an active area of research~\cite{fratus2022describing}.
As NMR spectroscopy plays a pivotal role in a wide range of industries from material chemistry~\cite{Bluemich2018,Reif2021} to medical related bio-medicine~\cite{molecules25204597,metabo12080678},
it is a perfect example to explore the capabilities of NISQ devices.
In this work, we fill the gap and provide insight into the noise tolerance of the digital quantum simulations of NMR experiments with NISQ devices. A key goal here is also the use of publicly availble devices that are quite easily available to the broader community. This will open the path to further improvements that might achieve practical applications in NMR spectroscopy.

\begin{figure}[t!]
    \centering
    \includegraphics[width=\columnwidth]{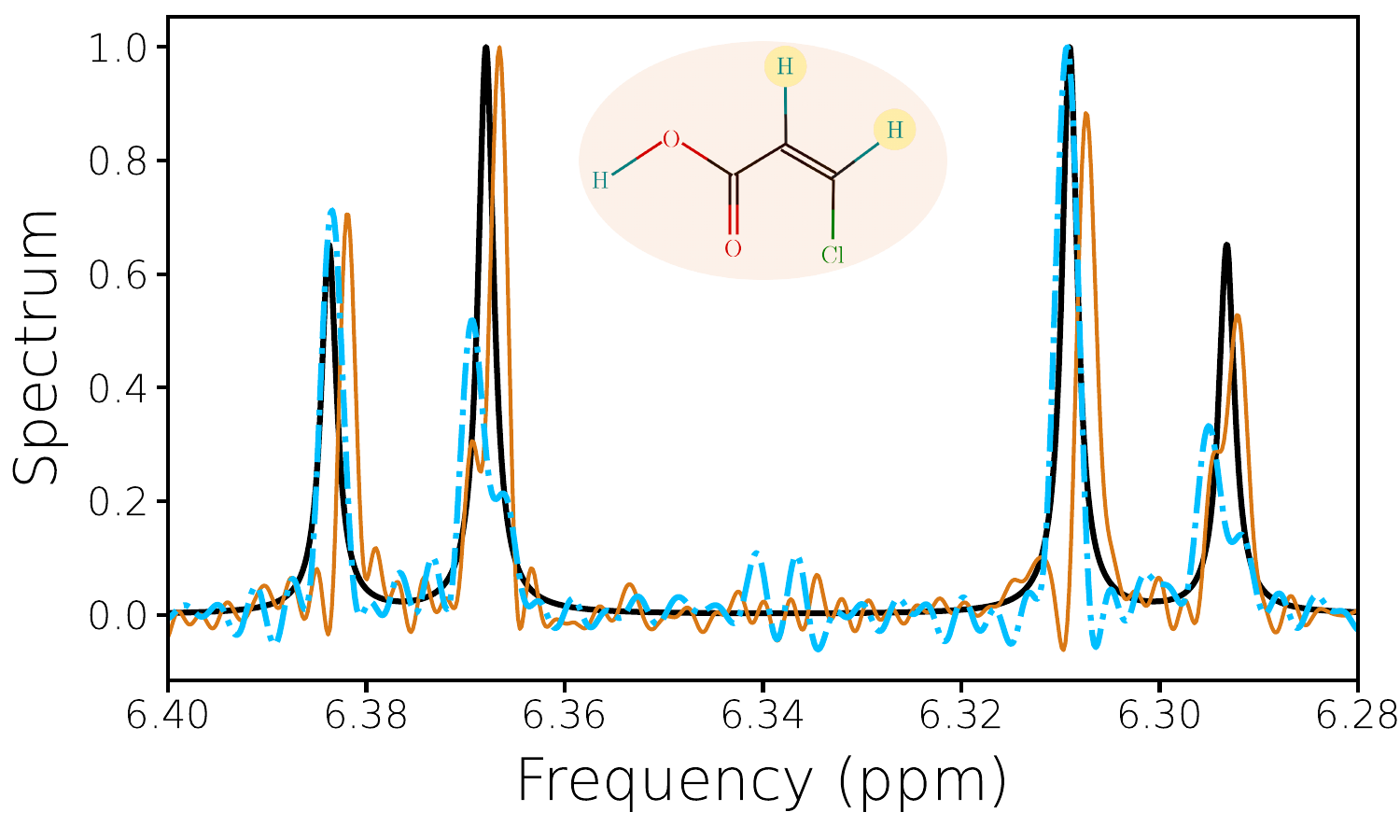}
    \caption{NMR spectrum of the non-exchangeable protons of \textit{cis}-3-chloroacrylic acid simulated on NISQ devices. The solid orange curve depicts the simulation using IBM Perth, the dash-dotted blue curve is the simulation using IonQ Aria via Amazon Braket, and the solid black curve shows the results from exact diagonalization. We have performed a time evolution for the spin Hamiltonian Eq.~(\ref{eq:2spin_ham}) at $B=11.7~T$ with Trotter step-size of $\tau=0.01$ and 81 and 61 Trotter-steps for IBM Perth and IonQ Aria via Amazon Braket, respectively. 
    }
    \label{fig:Spectrum_2spin_comparison_runs}
\end{figure}

To this end, we present the simulation of NMR spectroscopy of small organic molecules on publicly available quantum processors,
such as IBM's quantum processors based on superconducting qubits and IonQ's Aria trapped ion quantum computer available via Amazon Braket,
as it is showcased in Fig.~\ref{fig:Spectrum_2spin_comparison_runs} for NMR  spectrum of \textit{cis}-3-chloroacrylic acid.
The simulation on quantum computers nicely reproduce the known spectrum. However, even for such a small molecule [see Eq.~(\ref{eq:2spin_ham})],
we observe the signatures of noise as broadening of the predicted resonances.
This motivates further investigation of the impact of noise on quantum simulation of NMR spectroscopy, which is the main objective of the present work.
In the actual NMR experiments, the nuclear spins unavoidably interact with their environment, which is similar to the noise acting on qubits.
Therefore simulation of NMR spectra on quantum computers can endure a certain amount of noise.
We quantify this amount of noise and try to find a threshold for the noise that can be tolerated.


\section{The Spin Hamiltonian}
The NMR problem involves the interaction of the nuclear spins of molecules with an external magnetic field.
To tackle such a problem, we first express the molecules of interest in terms of spin-$1/2$ models that can be naturally mapped to qubits.
In the presence of a static magnetic field $\vec{B}$, the spin Hamiltonian of a given molecule in the liquid phase reads
\begin{equation}
H = -\gamma \sum_{i=1}^N (1+\delta_i) {\vec{S}_{i}} \cdot \vec{B} +  \sum_{i,j=1}^N\sum_{j>i} J_{ij}  \vec{S}_{i} \cdot \vec{S}_{j} ~,
\label{eq:Hamiltonian}
\end{equation}
where $\gamma$ is the gyromagnetic ratio, $N$ the total number of spins, $\delta_i$ encodes the chemical shifts caused by electrons shielding the magnetic field,
$\vec{S}$ represents the nuclear spin, and $J_{ij}$ determines the strength of the interaction between $i$-th and $j$-th nuclear spins. Note that in the liquid phase, the Hamiltonian has the SU(2) symmetry.
The chemical shifts and the spin–spin couplings are obtained through quantum chemical calculations (e.g., DFT calculations), similarly to the approach outlined in Refs.~\cite{Grimme2017,Yesiltepe2018,Willoughby2014,Jonas2022}.

In the present work, we focus on \textsuperscript{1}H NMR spectroscopy of two molecules, namely \textit{cis}-3-chloroacrylic acid ($C_3H_3O_2Cl$), and 1,2,4-trichlorobenzene ($C_6H_3Cl_3$). Both molecules have three spin-$1/2$ nuclei (protons).
However, due to the rapid exchange of the acidic proton in the solvent $D_2O$, the carboxylic signal is usually omitted. Therefore, the \textit{cis}-3-chloroacrylic acid is characterized by the following chemical shifts (in units of ppm) and spin-spin interaction (in Hz)
\begin{equation}
\delta=
\begin{pmatrix}
6.375 & 6.302
\end{pmatrix}
~,~
J_{12}=7.92~.
\label{eq:2spin_ham}
\end{equation}
The Hamiltonian of 1,2,4-trichlorobenzene ($C_6H_3Cl_3$) is described by
\begin{alignat}{1}
&\delta=
\begin{pmatrix}
7.194 & 7.377 & 7.467
\end{pmatrix},\\
&J_{1 2}= 8.5, J_{13}= 2.5, J_{23}= 0.5~.
\label{eq:3spin_ham}
\end{alignat}
The chemical shifts and the spin-spin couplings were determined to fit the experimental spectrum~\cite{NMR_data}. In addition, for 1,2,4-trichlorobenzene, the spin-spin couplings for the two opposing protons were taken from Ref.~\cite{NMR_par}.


\section{Method}
\label{Sec:Method}
In a 1D NMR experiment, an oscillatory external probe
is applied perpendicular to the static magnetic field mentioned above.
From now on, we assume that the static magnetic field ($B=11.7~T$) is applied in the $x$ direction.
Such a probe polarizes the nuclear spins and induces a time-dependent magnetic moment that can be expressed in terms of
correlation function \cite{fratus2025quantumcomputersimulatenuclear}
$A(t)=\text{Tr}\{S^{z}_{\rm tot}(t)S^z_{\rm tot}\}$
where $S^z_{\rm tot}=\sum_{i}S^z_{i}$ is the total nuclear spin that evolves in time as
\begin{equation}
S^{z}_{\rm tot}(t)=  e^{iHt}S^{z}_{\rm tot} e^{-iHt} .
\end{equation}

The NMR spectrum can be expressed in terms of the Fourier transform of the mentioned correlation function
\begin{equation}
A(\omega)=\text{Re}\left\{\int_{0}^{\infty}~dt~e^{-i\omega t-\Gamma t} \text{Tr}\{S^{z}_{\rm tot}(t)S^z_{\rm tot}\} \right\},
\end{equation}
where $\Gamma$ is the effective decoherence rate that physically arises due to the interaction of nuclear spin with its environment.

\begin{figure*}[t]
    \centering
    \includegraphics[width=0.9\linewidth]{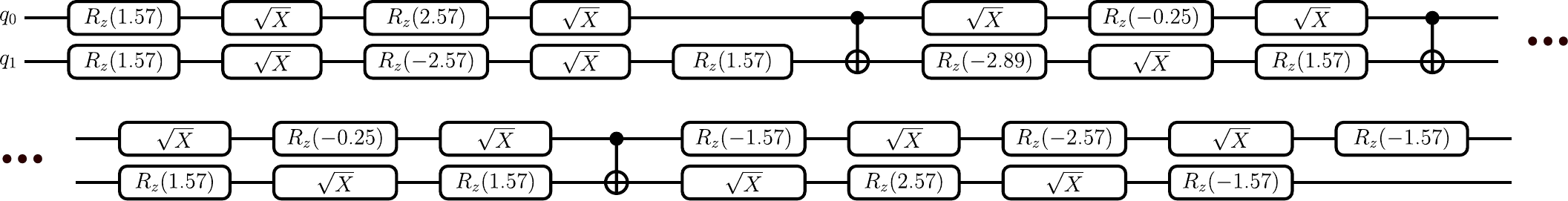}
    \caption{Circuit corresponding to one Trotter step ($\tau=0.01$) of the Trotterized time evolution of the
    Hamiltonian Eq.~(\ref{eq:2spin_ham}) for IBM devices with native gates CNOT, $X$, $\sqrt{X}$, and virtual $R_z$.
    For the 2-spin molecule, the interaction can be decomposed into three CNOT gates in the second-order Trotter expansion.}
    \label{fig:circuit_2_spin}
\end{figure*}

In order to find an algorithm for obtaining such a correlation function on a quantum processor,
we consider a Trotterized time evolution~\cite{Trotter1959, Childs2021},
where
the Hamiltonian of the NMR problem is written as the sum of partial Hamiltonians $\hat H=\sum_k \hat H_k$ and the time evolution is implemented according to
the Trotterization formula
\begin{align}
e^{-\textrm{i} \hat H t} \approx \left[ \prod_{k} e^{-\textrm{i} \hat H_k \tau} \right]^N  \, ,
\end{align}
with $\tau$ being the Trotter-step size and the approximation is controlled by the size of $\tau$ and $N$~\cite{Childs2021}.
For the NMR problem, the $\hat H_k$ terms correspond to the onsite energy terms and the interaction terms between spins.
The time propagation under a partial Hamiltonian can be implemented on the quantum computer directly,
using the unitary gates~$\hat U$ on the quantum computer.
\begin{align}
e^{-\textrm{i} \hat H_k \tau} = \prod_l \hat U_{k l}  \, .
\label{eq:evolution_gates}
\end{align}
Figure \ref{fig:circuit_2_spin} illustrates the quantum circuit corresponding to the Trotterized time evolution of the Hamiltonian in Eq.~(\ref{eq:2spin_ham}).

In a Trotterized time evolution,
the Trotter time step should be, on the one hand,
small enough to minimize the Trotter error and adequately cover the entire spectrum of the molecule.
On the other hand, the step size must be large enough to reduce the total number of Trotter steps required to simulate a sufficient length of the
time evolution. As we will see later, the size of the trotter step is inversely related to the effective width of the peaks in the simulated NMR spectrum.
In this respect, it is advantageous to perform the unitary transformation $H\to H-B\sum_{i}S^x_i$ that transforms the total magnetization in $z$-direction as
\begin{align}
e^{i tBS^{x}_{\rm tot}}S^{z}_{tot}e^{-i tBS^{x}_{\rm tot}}
&=\cos( B t)S^z_{\rm tot}+\sin( B t)S^y_{\rm tot}.
\end{align}
Consequently, we get (after neglecting counter-rotating terms)
\begin{align}
A(\omega)=\text{Re}\left\{\int~dt~e^{-i(\omega-B)t-\Gamma t} \text{Tr}\{ S^z_{\rm tot}(t)S^z_{\rm tot}(0)\} \right\}\nonumber\\
-\text{Im}\left\{\int~dt~e^{-i(\omega-B)t - \Gamma t} \text{Tr}\{ S^y_{\rm tot}(t)S^z_{\rm tot}(0)\} \right\}.\label{eq:spectral_function}
\end{align}
Furthermore, by introducing the eigenstates of $S^z_{\rm tot}(t)$ as following,
\begin{equation}
S^z_{\rm tot}(t)| m_n (t)\rangle=m_n (t)| m_n (t)\rangle~,
\end{equation}
the correlation functions can be rewritten according to
\begin{equation}
\text{Tr}\{ S^{z/y}_{\rm tot}(t)S^z_{\rm tot}(0)\} = 2\sum_{m^0_n>0} m^{0}_n \langle m_n(t)|S^{z/y}_{\rm tot}| m_n(t)\rangle ~,
\end{equation}
with $m_n^0=m_n(t=0)$~\cite{Seetharam_2023,Sels2020}. Note the sum can be restricted to positive initial magnetization $m_n^0>0$ due to the symmetry of the Hamiltonian
(which can be seen for example by flipping all spins around their X-axes, which changes  $S^{y/z}_i\rightarrow -S^{y/z}_i$,
but keeps the Hamiltonian unchanged).
In order to evaluate each term in the summation, we can initialize the qubits in a configuration leading to $m_n^0>0$,
and perform a Trotterized time evolution to obtain $\langle m_n(t)|S^{z/y}_{\rm tot}| m_n(t)\rangle$.

\begin{figure}[b!]
    \centering
    \includegraphics[width=\columnwidth]{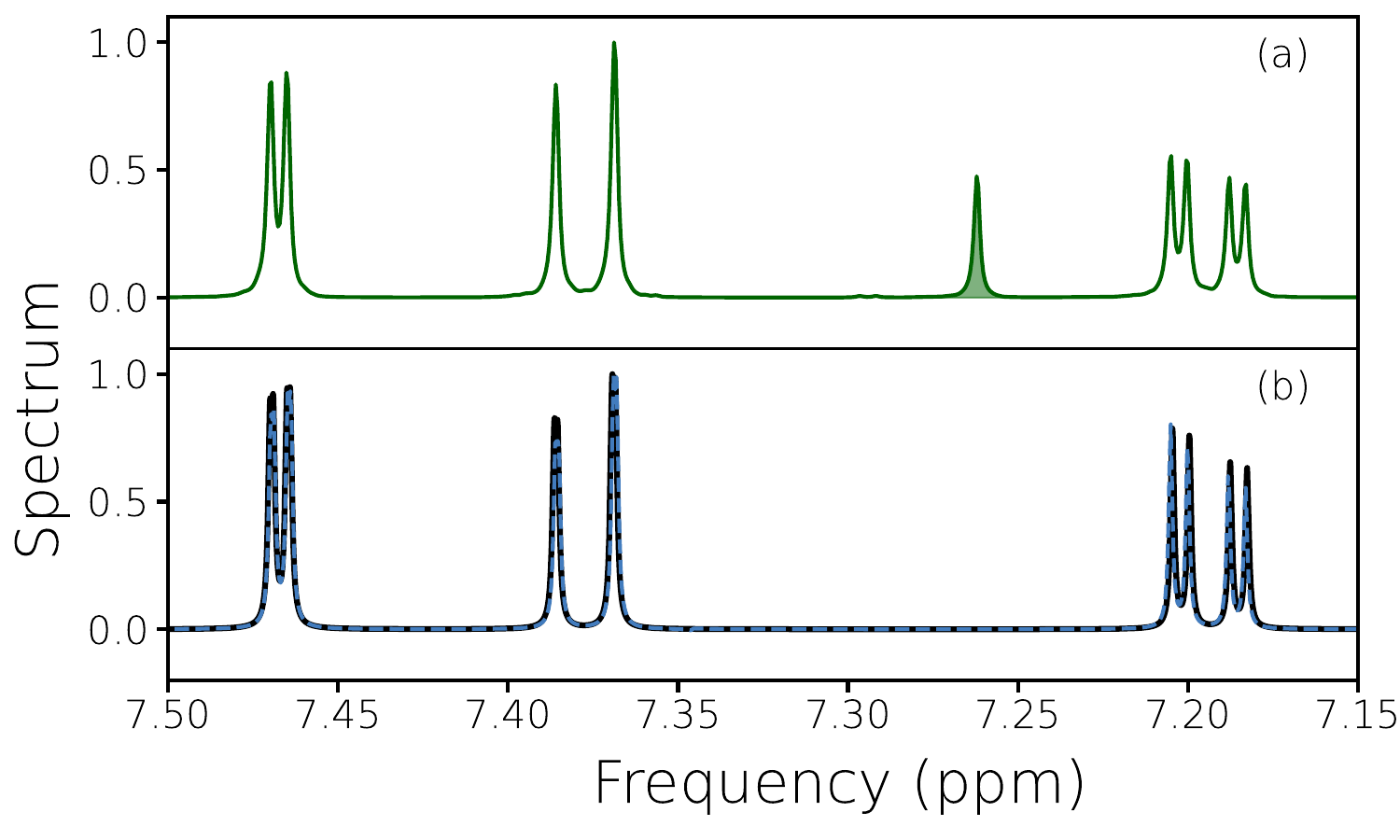}
    \caption{(a) Spectrum of 1,2,4-trichlorobenzene ($C_6H_3Cl_3$) from experimental NMR spectroscopy~\cite{NMR_data}. The marked peak at approx.~7.26 ppm is known to be caused by the solvent. 
    (b) Comparison of the simulation with exact diagonalization, shown by the solid black curve, with noise-free simulation illustrated by the dashed blue curve. For the noise-free simulation, we have used the algorithm presented in the Method section with 1000 Trotter steps of the size $\tau=0.005$.}
    \label{fig:fig0}
\end{figure}

In order to verify the outlined algorithm, we use a noise-free simulator to obtain the NMR spectrum of 1,2,4-trichlorobenzene ($C_6H_3Cl_3$) molecule and we compare it to the known experimental results~\cite{NMR_data}.
We choose the decoherence rate $\Gamma$, such that the simulated broadening of the peaks matches approximately the experimental broadening. As is shown in  Fig.~\ref{fig:fig0}, the simulated results agree very well with the experimental data, apart from the marked peak that is known to be caused by the solvent in which the sample is embedded.

\section{Results on NISQ devices}

In the following section, we present the outcome of executing the aforementioned method for organic molecules on quantum computing platforms.
In the 1D NMR experiment, the NMR response is measured at sufficiently long times after the applied oscillatory magnetic field,
such that the spin dynamics reach a steady state, in which the average magnetization is zero.
However, in our runs on current quantum computing platforms,
we were restricted with regard to the total depth of the circuits, i.e., the number of Trotter steps for the time evolution.
We have observed that such a maximum number of Trotter steps is smaller than the spin relaxation times,
and hence the oscillations in the averaged magnetization did not necessarily vanish at the end of the simulation, see Appendix \ref{sec:appendixA}.
Such a restriction also poses a maximum frequency grid for the spectrum.
In order to be able to obtain the spectrum more finely, we have added additional zeros in the averaged magnetization.


The spectrum of \textit{cis}-3-chloroacrylic acid is symmetric around the center of the rotating frame,
which is a manifestation of the symmetry of the interaction matrix of any two spin systems.
In general, the asymmetric part of the spectral function originates from the total magnetization in the $y$ direction, see Eq.~(\ref{eq:spectral_function}), which in the case of 2 spins vanishes.
In particular, the contributions from various spins should cancel each other.
However such a cancellation does not occur in the simulation results as noise manifests itself in disordered terms in the effective Hamiltonian being simulated on a noisy device, see Ref.~\cite{Reiner2018}.
In fact, if we impose such a symmetry,
and neglect the second contribution to Eq.~(\ref{eq:spectral_function}), we can significantly improve the results particularly for the IonQ device, see Appendix \ref{sec:appendixB}.

\begin{figure}[b!]
    \centering
    \includegraphics[width=\columnwidth]{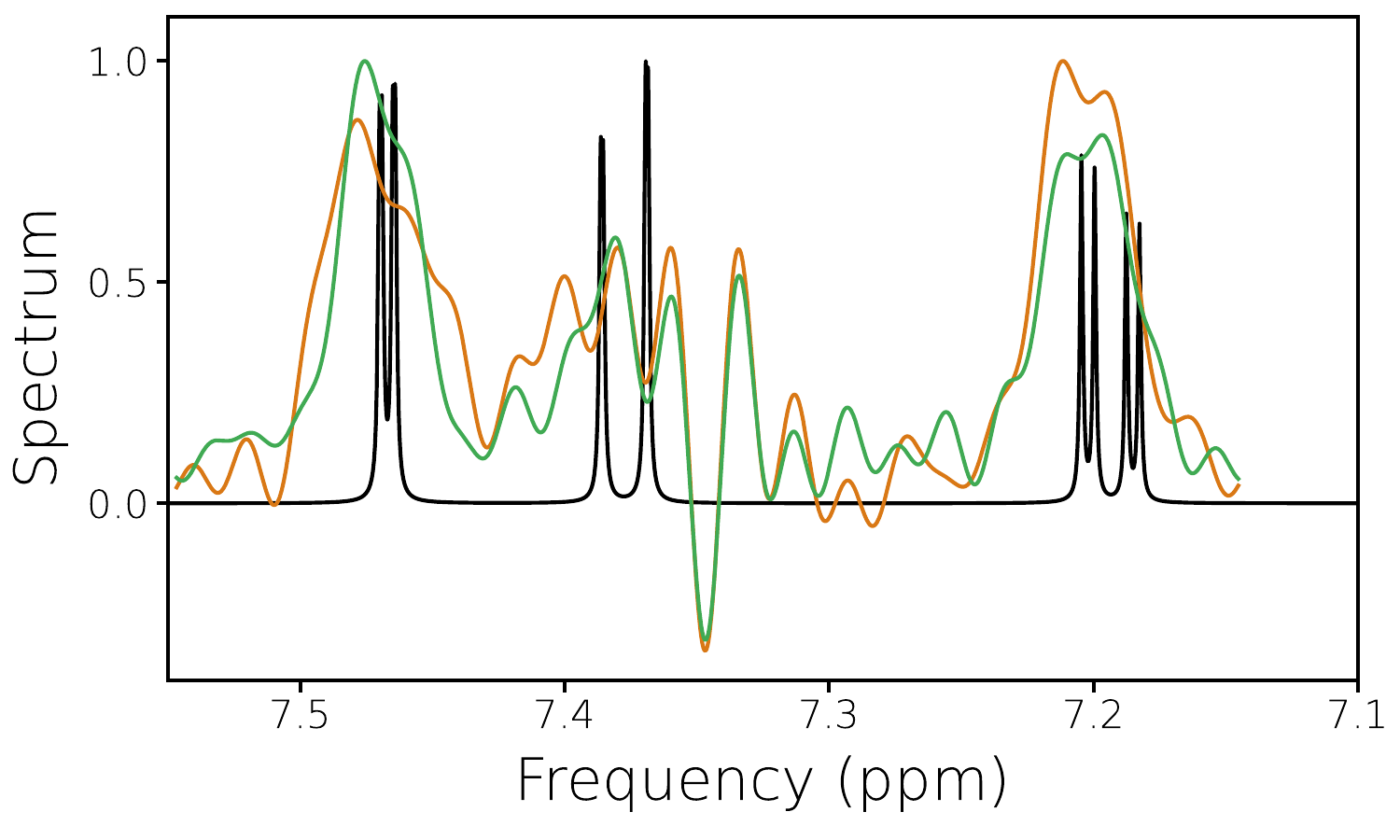}
    \caption{Comparison of the NMR spectrum of 1,2,4-trichlorobenzene ($C_6H_3Cl_3$) simulated on IBM Perth on two different days, depicted as the orange and green curves, with the exact results shown as the black curve. For the simulations on IBM, we have performed a time evolution with 21 Trotter steps of the size $\tau=0.005$.}
    \label{fig:Spectrum_3spin_comparison_runs}
\end{figure}

Figure \ref{fig:Spectrum_3spin_comparison_runs} depicts the spectrum of 1,2,4-trichlorobenzene simulated using the IBM Perth devise on two different days.
As illustrated, the impact of noise on the simulations is so significant that we cannot reproduce all the fine structures of the eaxct NMR spectrum.
This is due to the fact that the Hamiltonian of 1,2,4-trichlorobenzene has three spins with all-to-all interactions, see Eq.~(\ref{eq:3spin_ham}),
On the IBM-Q device, we selected a three-qubit chain to execute the algorithm.
Due to the Hamiltonian's all-to-all interaction, it is necessary to swap the positions of the qubits.
For simulating NMR spectra of 1,2,4-trichlorobenzene, we employed a second-order Trotter expansion.
This approach results in application of 15~CNOT gates during each Trotter step.
Moreover, due to this large-depth circuit,
we have picked a smaller number of Trotter steps. This
number is far below the required time for the average
magnetization to reach zero, and hence the added zeros
(see the above paragraph) cause an abrupt change in the
magnetization. This, in turn, results in the oscillations
around the center of the rotating frame.

\section{Noise analysis}
\begin{figure}[t!]
    \centering
    \includegraphics[width=\columnwidth]{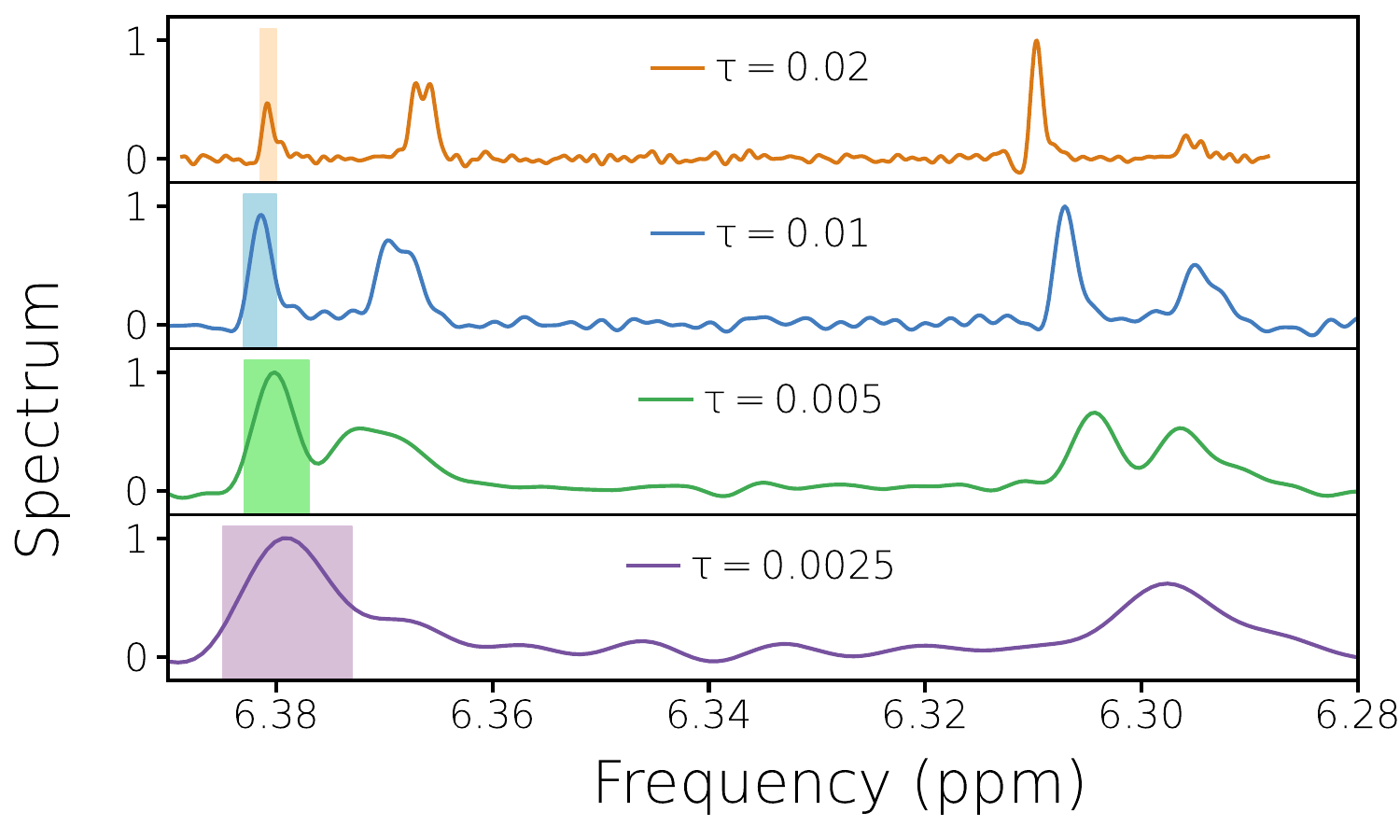}
    \caption{Comparison of the NMR spectrum of \textit{cis}-3-chloroacrylic acid simulated on IBM Perth device for a varying Trotter step size, with fixed number of Trotter steps (81). The shaded area depicts the effective decoherence rate Eq.~(\ref{eq:effective_decoherence_rate}) in ppm units.}
    \label{fig:noise_scaling}
\end{figure}

A critical component of every execution of any algorithm on current quantum computers is the effect of noise.
For any algorithm, it would be of great use to have a way to construct estimates for the quality of the execution on a NISQ computer
purely from the noise characteristics of the quantum gates. Normally, the noise characteristics are characterized by the quantum computing providers
in terms of error probabilities $\epsilon_{kl}$ or gate fidelities $F_{kl}=1-\epsilon_{kl}$, where every gate in the implementation of a Trotterized time evolution, see Eq.~\ref{eq:evolution_gates}, is associated with a Fidelity and error probability. We will now discuss why the width of the peaks in the simulated NMR spectra can be roughly estimated to be
 \begin{equation}
\Gamma_{\rm eff}=\frac{\sum_{kl}\epsilon_{kl}}{N\tau},
\label{eq:effective_decoherence_rate}
\end{equation}
where $\epsilon_{kl}$ is the error probability of each gate that is applied in the algorithm, see Eq.~\ref{eq:evolution_gates} and Fig.~\ref{fig:circuit_2_spin}, where $N$ is the number of qubits in the simulation and $\tau$ is the Trotter time step as discussed above. In Fig.~\ref{fig:noise_scaling}, we illustrate the simulated NMR spectrum on the IBM quantum computer for various Trotter steps. The figure also shows the estimate of the peak width $\Gamma_{\rm eff}$ which was purely done using data provided by IBM, and without any additional fitting. The results indicates that the scaling on the trotter step size behaves exactly as predicted by Eq.~\ref{eq:effective_decoherence_rate}.

The inverse proportion 
of the peak width with the Trotter time step is in itself a major result of this work. In principle, it is clear that the noise will simply scale linearly with the number of applied gates and the number of applied gates scales linearly with the Trotter time step if total simulation time is kept constant. But it is often not appreciated that this trade off exists and a larger Trotter step might be preferable (due to suppression of the impact of noise) despite the increased Trotter error.

\begin{figure}[t]
    \centering
    \includegraphics[width=\columnwidth]{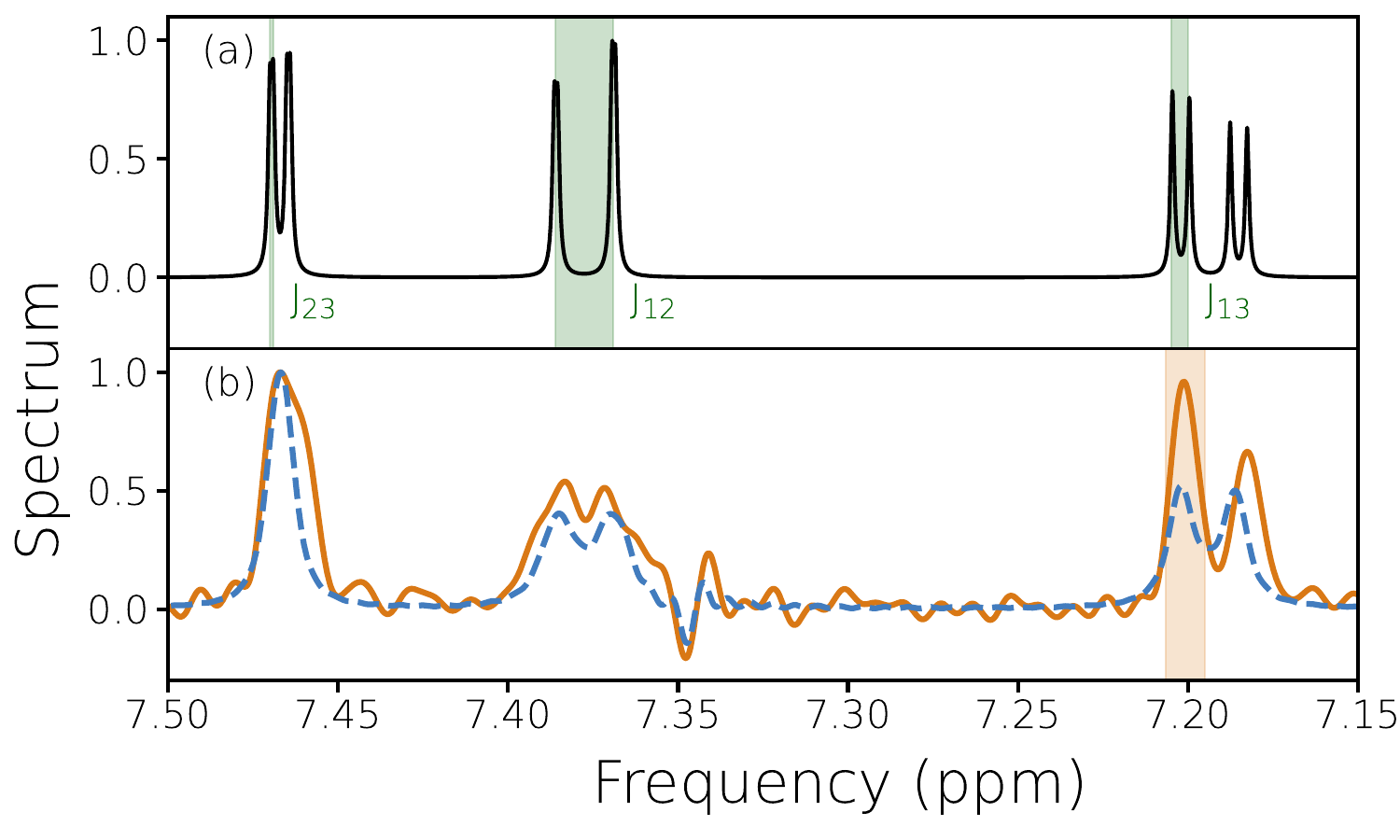}
    \caption{(a) NMR spectrum of  1,2,4-trichlorobenzene ($C_6H_3Cl_3$) obtained via exact diagonalization. The shaded area depicts the different spin couplings in ppm units, see Eq.~\ref{eq:3spin_ham}. (b) Comparison of the NMR spectrum of the reduced Hamiltonian of 1,2,4-trichlorobenzene ($C_6H_3Cl_3$) simulated on IBM Perth (solid orange curve) with on-gate depolarizing noise model emulation (dashed blue curve). The shaded area depicts the effective decoherence rate Eq.~\ref{eq:effective_decoherence_rate} in ppm units.
    }
    \label{fig:spectrum_ham_reduced}
\end{figure}

In Fig. \ref{fig:spectrum_ham_reduced}, we show the linewdith given by $\Gamma_{\rm eff}$ as compared to an actual run on a quantum computer for a three spin example, see Eq.~\ref{eq:3spin_ham}. We see that there is again a good agreement. This example is quite important and interesting to understand how we can directly use this simple rate to make meaningful predictions for the results of our calculation and even use it to improve our calculations. As is clear from the fact that
$\Gamma_{\rm eff}$ is an effective line width, it represents the limit of energy scales that can be resolved. In fig. \ref{fig:spectrum_ham_reduced}.~(a), we show the coupling $J_{23}$ and in fig. \ref{fig:spectrum_ham_reduced}.~(b) we see that the size of $\Gamma_{\rm eff}$ is orders of magnitudes larger than the spin coupling,  $\Gamma_{\rm eff}>J_{23}$. Note that all this is known already before actually simulating the spectra. The coupling known from the input Hamiltonian and the rate $\Gamma_{\rm eff}$ can be calculated using the parameters provided by the quantum computing providers. Therefore we can directly simplify the problem by neglecting $J_{23}$ which reduces the gates depth of the circuit.
Neglecting the coupling $J_{23}$ simplifies 1,2,4-trichlorobenzene to a nearest neighbor Hamiltonian.
In the first-order Trotter expansion, each Trotter circuit then contains 6 CNOT gates (instead of 15 CNOTs).
As a result, the simulated NMR spectrum of the reduced Hamiltonian is significantly improved,
as can be seen by comparing Fig.~\ref{fig:Spectrum_3spin_comparison_runs} with Fig.~\ref{fig:spectrum_ham_reduced}.
For completeness, we have also presented the simulation with depolarising noise, and we see that the resulting spectrum agrees well with the simulated one on IBMQ cloud.
\subsection*{Deriving the effective line width}

Having seen that the proposed effective linewidth Eq.~\ref{eq:effective_decoherence_rate} agrees well with the simulated spectrums on IBM's quantum computers,  we now outline how to obtain such a linewidth $\Gamma_{\rm eff}$. This derivation is based on the assumption that the error probability $\epsilon_{kl}$ is caused by incoherent errors. As shown above this seems to be a good approximation, but it is also possible to turn coherent error into incoherent error using randomized compiling without gate overhead\cite{Hashim2021} and hence the applicability of our assumption can be generalized.

Typically each single qubit is characterized by the parameters $T_1$ and $T_2^*$~\cite{Klimov_2018,Schl_r_2019}. The former is the damping time scale, i.e., the time scale after which a qubit in an excited state will decay to its ground state.The second time scale $T_2^*$ is pure dephasing time, which refers to the time scale, after which the qubit's phase is lost (when removing the contribution from damping).
However, it has been shown in Ref.~\cite{fratus2023discrete}, that large angle gates drastically modify the form of such noise processes, leading to a form close to depolarizing noise~\cite{fratus2023discrete}.
And hence as all the native gates for IBM's quantum processors are corresponding to large angle rotations (except the $R_{z}$ which is a virtual gate),
in our modeling of  IBM's quantum computing devices, we model the gate errors as depolarizing noise.
This noise model is also consistent with noise spectroscopy performed on the IBM devices~\cite{Stadler2024}. However, even if noise would correspond to
decay or pure dephasing this would not fundamentally change the effective line width. Incoherent errors are in general described by
using a Lindblad superoperator $\mathcal{L}_{kl}$, with the noise strength characterized by decoherence rates $\gamma^{i,kl}_\textrm{depol}$
acting on qubit $i$. Since the error probabilities are given for each gate, it is best to model the effect of noise as a depolarising channel that acts after
each gate. That is why we used the indices $k$ and $l$ for the Lindblad superoperator and the decoherence rate since each of these operators is associated
with a unitary operations $\hat{U}_{kl}$.

To determine the decoherence rates of the depolarising noise after each gate $\hat{U}_{kl}$,
we use the universal connection between incoherent noise and gate fidelity~\cite{Abad2022},
which yields
\begin{equation}\label{eq:connect_error_to_noise}
\epsilon_{kl}=\frac{3d}{4(d+1)}t_{\rm gate,kl}\sum_i^{n}\gamma^{i,kl}_\textrm{depol},
\end{equation}
where $n$ is the number of qubits the gate is acting on, $d=2^n$ is the dimension of the space and $t_{\rm gate,kl}$ is the gate execution time.
The gate execution time is not necessarily exactly known, but as we will see it is also not necessarily needed. The relevant
part of Eq.~\ref{eq:connect_error_to_noise} is to show that there is a connection between error probabilities and decoherence rates. This is what we will use to derive $\Gamma_{\rm eff}$.

Since we think about a noisy system, we consider the time evolution of the
density matrix $\rho$ and we can write the application of one Trotter step as
\begin{equation}\label{eq:time_evolve_density}
 \rho(\tau_0+\tau)=\prod_l e^{\mathcal{L}_{kl}t_{\rm gate,kl}}\mathbf{U}_{kl}\rho(\tau_0)
\end{equation}
with $\mathbf{U}_{kl}\rho(\tau_0) =\hat{U}_{kl}\rho(\tau_0)\hat{U}_{kl}^{\dagger}$. It should be noted that we used the effective
time of the Trotterized time evolution $\tau$ to indicate the time evolution of the density matrix. However, the effect of the noise is of course
related to the physical gate execution time $t_{\rm gate,kl}$. For a Trotterized time  evolution it is possible to efficiently commute
all unitary operations in Eq.~\ref{eq:time_evolve_density} to the left and all incoherent noise terms to the right.
If we then rescale all depolarizing rates to
\begin{equation}
\tilde{\gamma}^{i,kl}_\textrm{depol}\tau=\gamma^{i,kl}_\textrm{depol}t_{\rm gate,kl}
\end{equation}
we can write the master equation as
\begin{equation}
 \rho(\tau_0+\tau)= e^{([H,\bullet]+\mathcal{L}_{\rm incoherent})\tau} \rho(\tau_0)
\end{equation}
where $e^{[H,\bullet]}$ represents the coherent time evolution we wanted to create using the Trotterization algorithm and
$e^{\mathcal{L}_{\rm incoherent}\tau}$ contains all incoherent effects of the depolarizing noise.
Details on the exact derivation of
$\mathcal{L}_{\rm incoherent}$ can be found in Ref.~\cite{fratus2022describing}. Additionally, simple examples for the form of $\mathcal{L}_{\rm incoherent}$
can be found in Appendix B of Ref.~\cite{Juha2023}. However, the relevant transformations change the structure of the noise operators,
but leaves the total sum of all incoherent processes $\sum_{ikl} \tilde{ \gamma}^{i,kl}_\textrm{depol}$ unchanged.
It should be noted that all depolarising rates
$\tilde{ \gamma}^{i,kl}_\textrm{depol}$
in $e^{\mathcal{L}_{\rm incoherent}\tau}$ are now connected to the error probabilities as
\begin{equation}
\sum_{kl} \epsilon_{kl}\approx\frac{1}{2}\tau\sum_{kl}\sum_i^{n}\tilde{\gamma}^{i,kl}_\textrm{depol},
\end{equation}
where we have simplified the prefactor in Eq.~\ref{eq:connect_error_to_noise} to $1/2$, which is a fair approximation if only
one- and two-qubit gates are involved. Now if we assume that either a SWAP algorithm is used
or that error probabilities are similar for all qubits used we get an effective averaged
single spin decoherence rate by dividing the total sum of all rates by the number of spins, resulting in the full-width-half-maximum of
\begin{equation}
 \Gamma_{\rm eff}=\frac{\sum_{kl}\epsilon_{kl}}{N\tau},
\end{equation}
which effectively is the sum over all incoherent rates divided by the number of qubits.
Such an effective decoherence rate can be also obtained from the equation of motion of correlation function within the effective open quantum system description, as is presented in Appendix \ref{sec:appendixE}.

\subsection*{Discrete noise model in quantum circuits}

The imperfections of the qubits typically results in incoherent errors, i.e., qubit damping~\cite{Klimov_2018}, dephasing~\cite{Schl_r_2019}, or depolarization after the application of gates~\cite{fratus2022describing,fratus2023discrete}.
Additionally, another source of error is the erroneous unitary operation implemented by the gates.
Such coherent error can be interpreted as effective additional unitary gate.
Moreover, leakage out of the computational basis of the qubit~\cite{Chen_2016}, and a fast return back to mentioned basis may be mapped to incoherent errors. In this section, we expand the noise model beyond incoherent errors. This allows us to reproduce most of the features of the shown spectrums simulated on IBM quantum processors.

\begin{figure}[t!]
    \centering
     \includegraphics[width=\columnwidth]{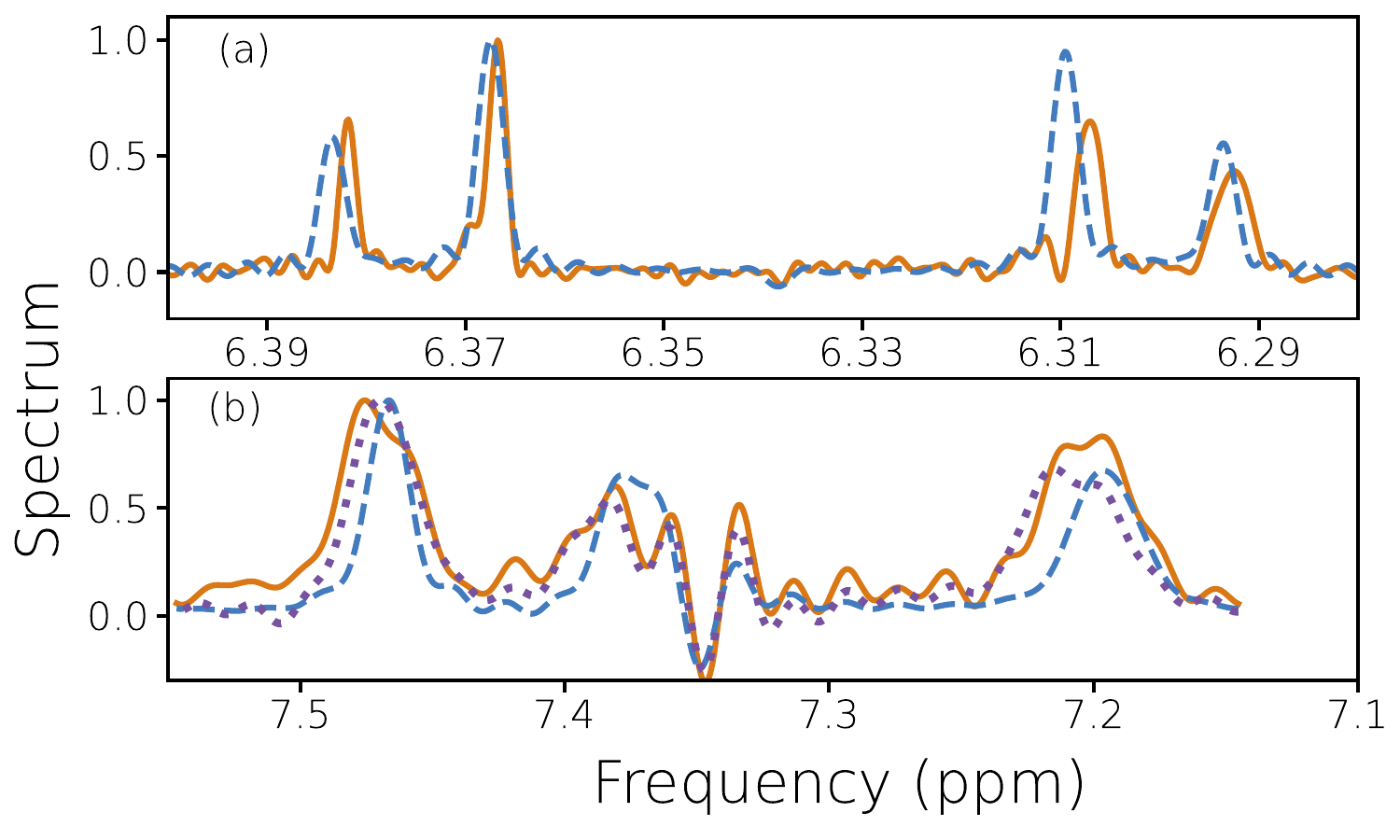}
    \caption{(a) Comparison of the NMR spectrum of \textit{cis}-3-chloroacrylic acid simulated on IBM Perth (solid orange curve) with on-gate noise model emulation (dashed blue curve) for 81 Trotter steps of size $\tau=0.01$.
    (b) Comparison of the NMR spectrum of 1,2,4-trichlorobenzene ($C_6H_3Cl_3$) simulated on IBM Perth,
    shown as orange solid curve, with both the numerical model emulation including only depolarising noise shown as the dashed blue curve,
    and a noise model including also coherent errors illustrated by the dotted magenta curve, for 21 Trotter steps of size $\tau=0.005$.
    In the latter noise model, we have assumed a coherent Z-error with angle $\phi_{z}=-0.027$, after each gate operation. After each two-qubit gate operation (here, CNOT operations),
    we have assumed a coherent error of the form $R_x$, with angle $\phi_{x}=0.05,-0.047$ on the control and target qubits, respectively, and an additional error $R_z(-0.05)$ solely on the control qubit.}
    \label{fig:spectrum_3spin_comparison_sim}
\end{figure}

Figure \ref{fig:spectrum_3spin_comparison_sim}(a)
illustrates the comparison of the simulation which includes only depolarising noise according to Eq.~(\ref{eq:connect_error_to_noise}) with the IBM results for \textit{cis}-3-chloroacrylic acid.
As it is shown, the agreement is rather good.
The peak positions show slight differences and also vary between different runs, see Appendix \ref{app:additional_results},
which is a manifestation of the effect of quasistatic coherent errors. 
Also the peak widths can vary between different runs.

Figure \ref{fig:spectrum_3spin_comparison_sim}(b)
shows the comparison of the simulation which includes only depolarising noise (blue dashed lines) with the IBM results for
the case of 1,2,4-trichlorobenzene. This system involves more interaction terms and hence more gates in the circuit for time evolution [in total 15 CNOTs per Trotter step,
in comparison to 3 CNOTs in case~(a)].
The latter results in more prominant impact of coherent errors and hence a considerable deviation from the simulation with including only depolarising noise.
As demonstrated, in the second numerical simulation (dotted lines),
where we include coherent errors by applying an additional $R_{z}$ and $R_{x}$ on each involved gate (determined by a fitting routine), we can better reproduce the results obtained on IBM devices by such an inclusion of coherent errors.
We use angles whose magnitudes are consistent with the noise spectroscopy of Ref.~\cite{Stadler2024} and correspond to gate errors that are smaller than given in the calibration data.
As illustrated, with the proper choice of angles corresponding to coherent errors, we are able to reproduce the main features of the simulated spectrum on IBM.
This demonstrates that also coherent errors can effectively broaden the spectrum in large-depth circuits.

\section{Conclusion}

We demonstrated the digital quantum simulation of NMR experiments,
employing publicly available quantum processors. We compared the results against conventional methods to study the effect of noise.
We provided a noise model that can accurately reproduce the main features of the obtained spectra.
We formulated an effective decoherence rate, based on the noise parametrization of the quantum devices,
which can define the energy scale up to which we can accurately simulate NMR spectra with a given NISQ device.
Moreover, we showcase how such an effective decoherence rate provides insight into tailoring the circuit to a smaller depth, and, in turn, improve the obtained NMR spectra.

In this work, we could only show the simulation of NMR spectra for very small organic molecules due to present limitations of NISQ devices.
However we witness a surge of quantum computing platforms that continue to proliferate with ever increasing qubit gate fidelities. With the increase of the quantity and quality of quantum computers, it remains quite relevant to find interesting applications.
In this respect, our results highlight a promising application for publicly available quantum computers,
as well as showcasing how one can tailor the quantum algorithms through noise analysis, paving the way for better employing quantum tasks on NISQ devices.

\begin{acknowledgments}
%
This work was supported by the German Federal Ministry of Education and Research, through projects QSolid (13N16155) and Q-Exa (13N16065).
We would like to thank Peter Pinski, Sonia Alvarez, Julius Kleine B{\"u}ning, and Jesse~A.~Vaitkus for insightful input and fruitful discussions.
\end{acknowledgments}
\bibliography{ref}


\appendix

\section{Time evolution of Magnetization}
\label{sec:appendixA}

\begin{figure}[t!]
    \centering
    \includegraphics[width=\columnwidth]{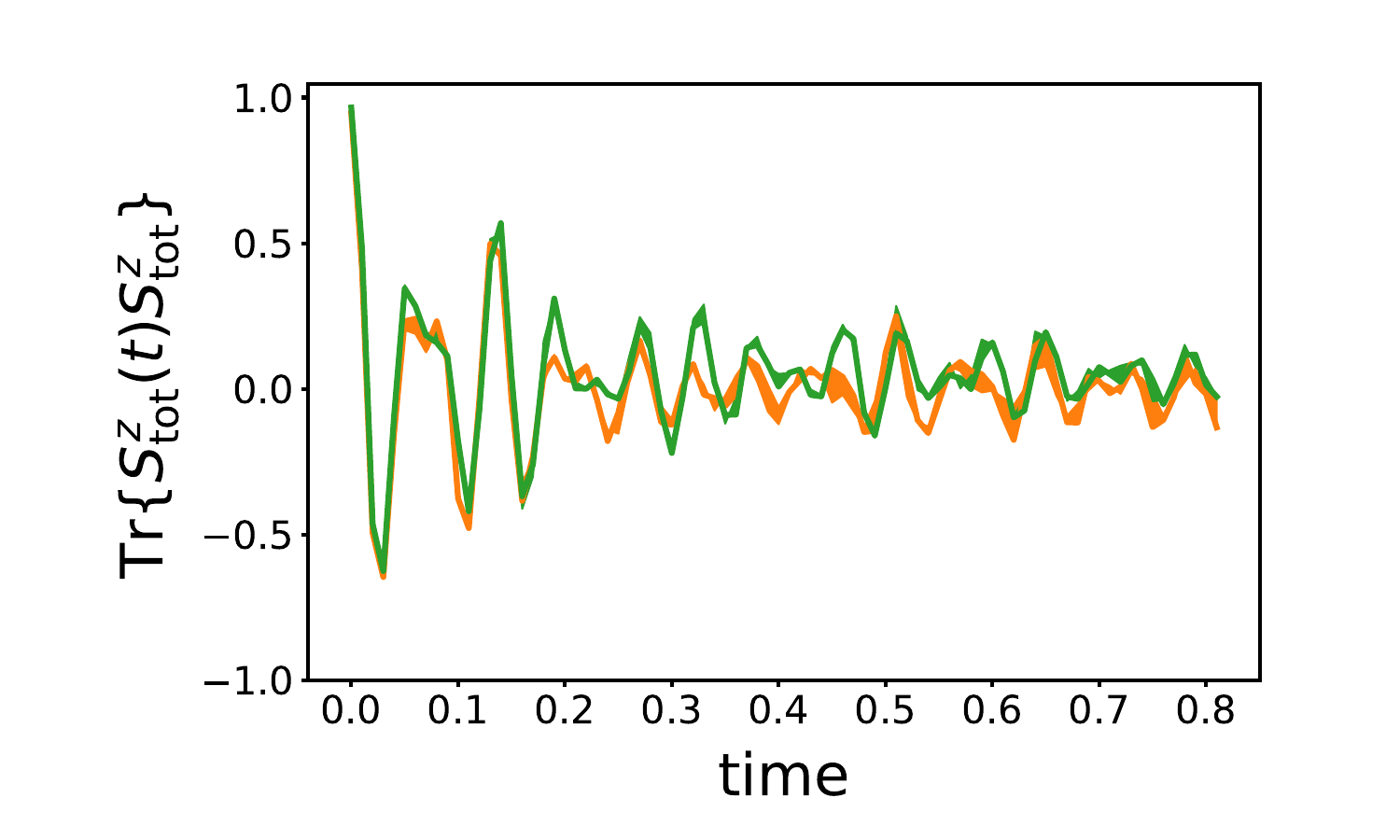}
    \caption{Time evolution of the correlation function of the \textit{cis}-3-chloroacrylic acid, simulated on two different devices from IBM quantum computers, namely IBM Perth (orange curve) and IBM Nairobi (green curve), with 81 Trotter steps of size $\tau=0.01$.}
    \label{fig:time_evolution_correlation_function_IBM}
\end{figure}

\begin{figure}[t!]
    \centering
    \includegraphics[width=\columnwidth]{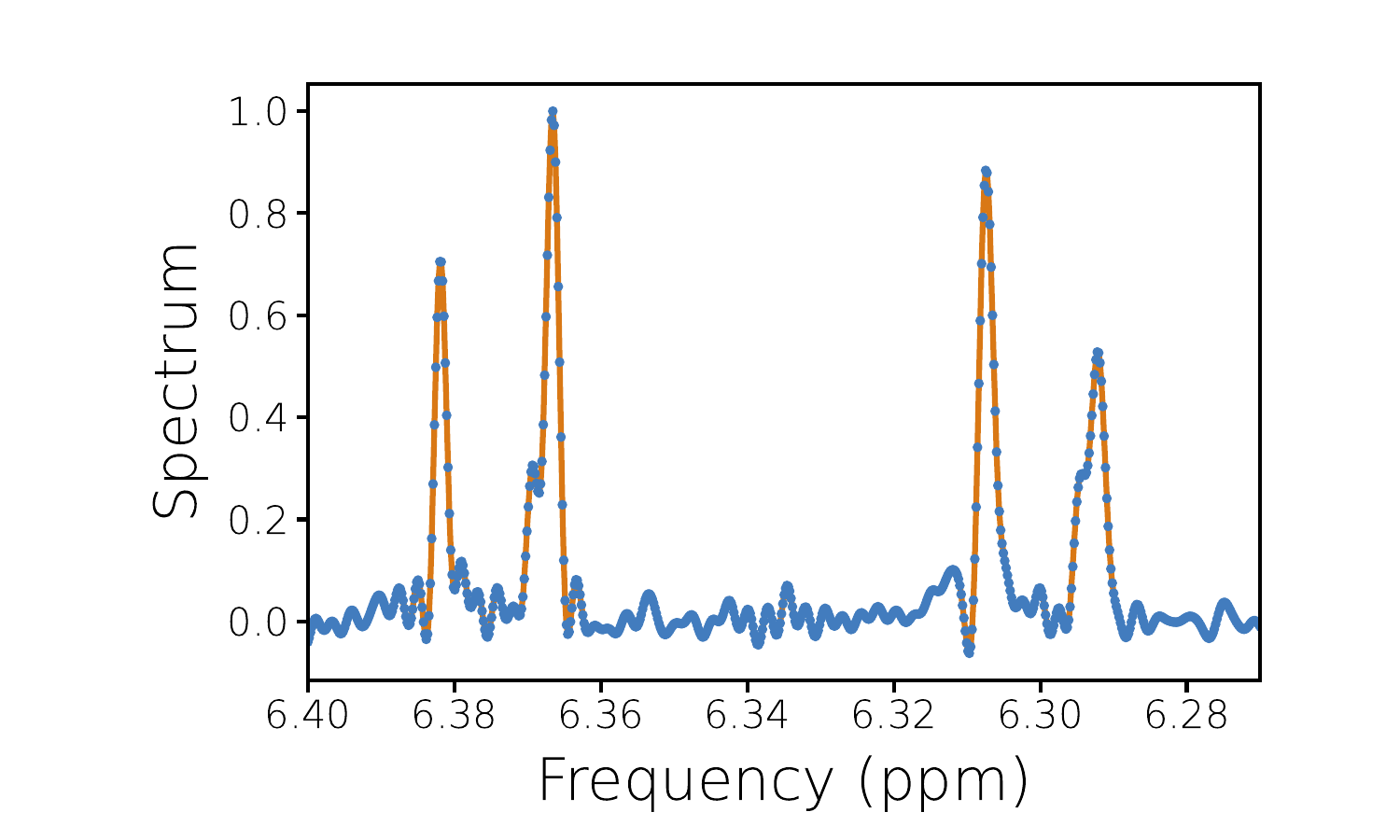}
    \caption{Comparison of the zero-padded spectrum of \textit{cis}-3-chloroacrylic acid simulated on IBM Perth illustrated by orange curve with original results shown as blue dots.}
    \label{fig:spectrum_added_zeros}
\end{figure}

In this section, we look at the time evolution of the correlation function simulated on IBM quantum computing platforms. Figure \ref{fig:time_evolution_correlation_function_IBM} depicts the correlation function $\text{Tr}\{ S^z_{\rm tot}(t) S^z_{\rm tot}\}$ as a function of time, indicating the decay of correlation function with time. In the NMR experiment the spectrum is measured at sufficiently long times after the application of the magnetic field such that the system reaches a state in which average magnetization vanishes and hence the correlation function.
As is shown in Fig.~\ref{fig:time_evolution_correlation_function_IBM} there remain some small oscillations around zero
at long times due to the limitation of number of Trotter steps.
As the spectrum is obtained via the Fourier transform of the correlation function, see Eq.~(\ref{eq:spectral_function}), the frequency interval is inversely proportional to the number of Trotter steps.
Thereby we added additional zeros to the correlation functions in order to be able to obtain the spectrum on a finer frequency grid.
Fig.~\ref{fig:spectrum_added_zeros} presents the comparison of the spectrum with and without adding such additional zeros for one of the simulations on IBM's devices.

\section{The symmetrized NMR spectra}
\label{sec:appendixB}

\label{sec:appendixB}
\begin{figure}
    \centering
    \includegraphics[width=\columnwidth]{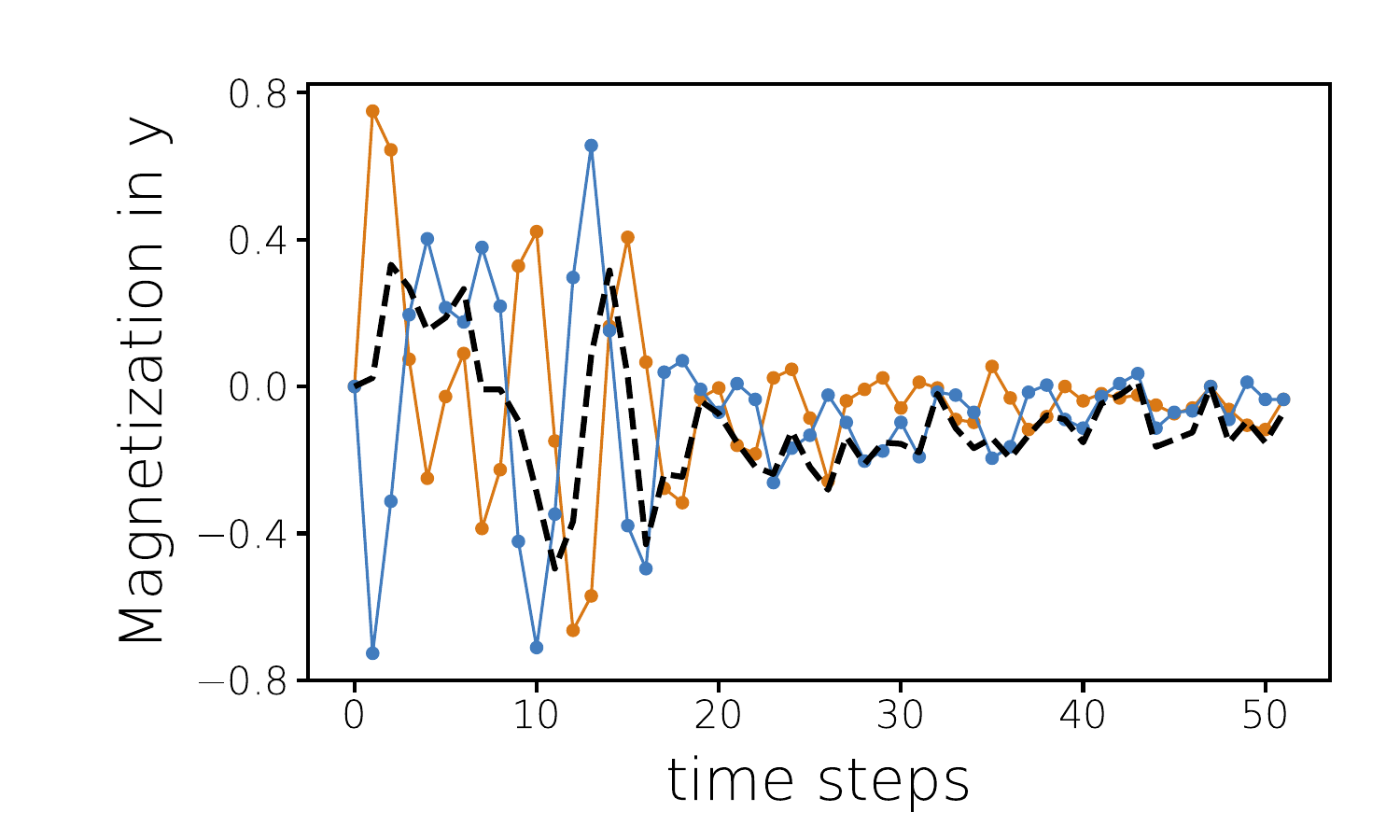}
    \caption{Time evolution of the magnetization in the $y$ direction for \textit{cis}-3-chloroacrylic acid with Hamiltonian Eq.~\ref{eq:2spin_ham}, simulated on IonQ Aria, with 61 Trotter steps of size $\tau=0.01$. The contribution from each spin to the magnetization (orange and blue curve) is compared with the total one (black curve).}
    \label{fig:magnetization_in_y}
\end{figure}

We discuss the symmetry breaking that occurs in the simulations as an artifact of noise of NISQ devices.
In particular in Fig.~\ref{fig:magnetization_in_y}, we present the magnetization in $y$ direction simulated on IonQ's Aria. Note that due to the symmetry of the Hamiltonian of  \textit{cis}-3-chloroacrylic acid, see Eq.~\ref{eq:2spin_ham}, we expect $\langle S^0_y \rangle=-\langle S^1_y \rangle$,  such that the total magnetization in the $y$ direction vanishes. However, due to the presence of noise, such a symmetry is numerically broken.
This is a manifestation of coherent errors~\cite{Reiner2018} resulting in additional disordered effective Hamiltonian terms that break the mentioned symmetry.
In particular, in additional numerical simulations we find that the spectroscopy on quantum computer is especially fragile against Hamiltonian disorder terms~$\propto YZ$.
\begin{figure}
    \centering
    \includegraphics[width=\columnwidth]{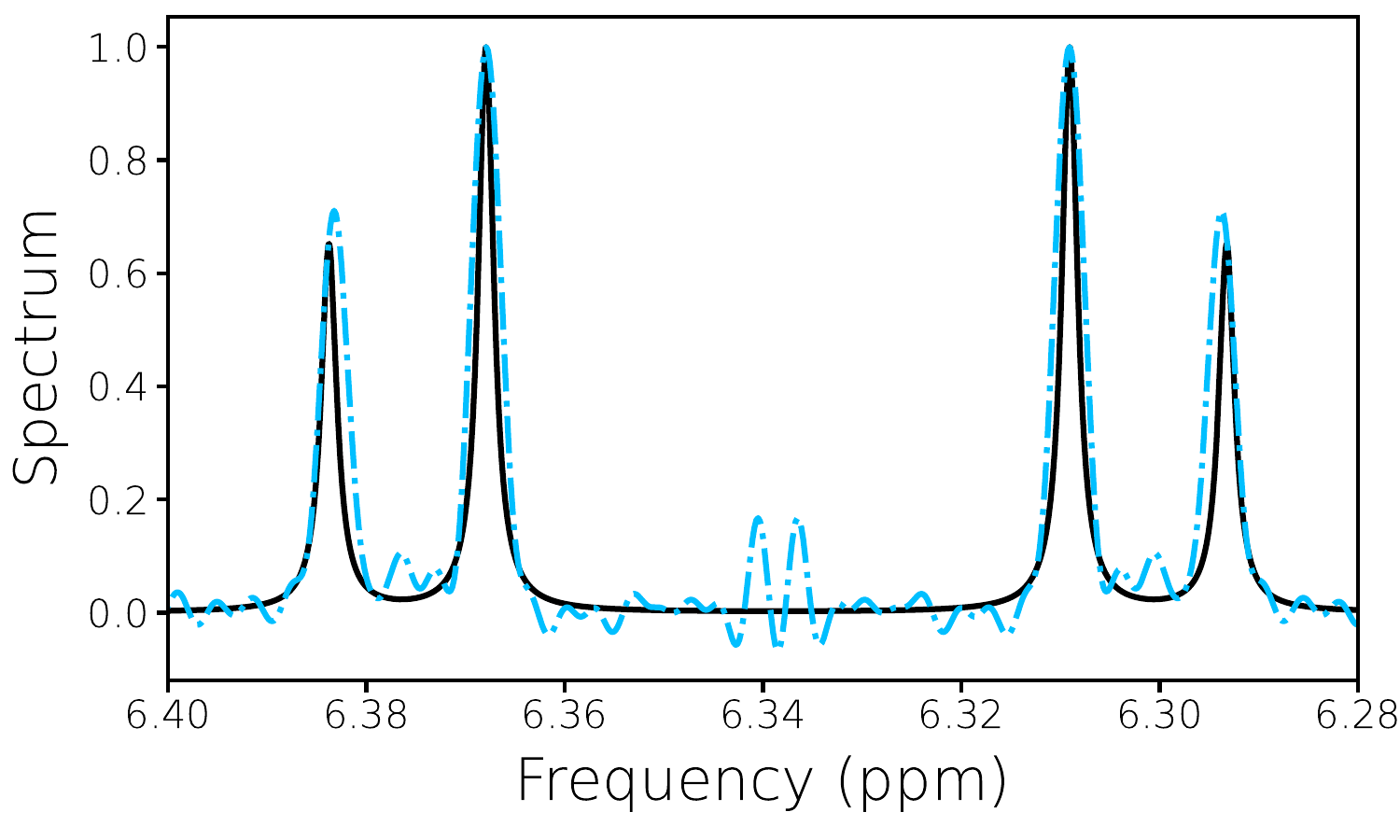}
    \caption{Comparison of the symmetrized spectrum of the \textit{cis}-3-chloroacrylic acid, simulated on IonQ Aria (dash-dotted blue curve) with 61 Trotter steps of size $\tau=0.01$, with exact diagonalization results (black solid curve).}
    \label{fig:spectrum_sym_IonQ}
\end{figure}
If we impose such a symmetry by neglecting the second term in Eq.~\ref{eq:spectral_function}, we can significantly improve the results as is shown in Fig.~\ref{fig:spectrum_sym_IonQ} for the simulation on IonQ Aria.

\section{The Trotter error}\label{app:trotter_error}
\label{sec:appendixC}
\begin{figure}
    \centering
    \includegraphics[width=\columnwidth]{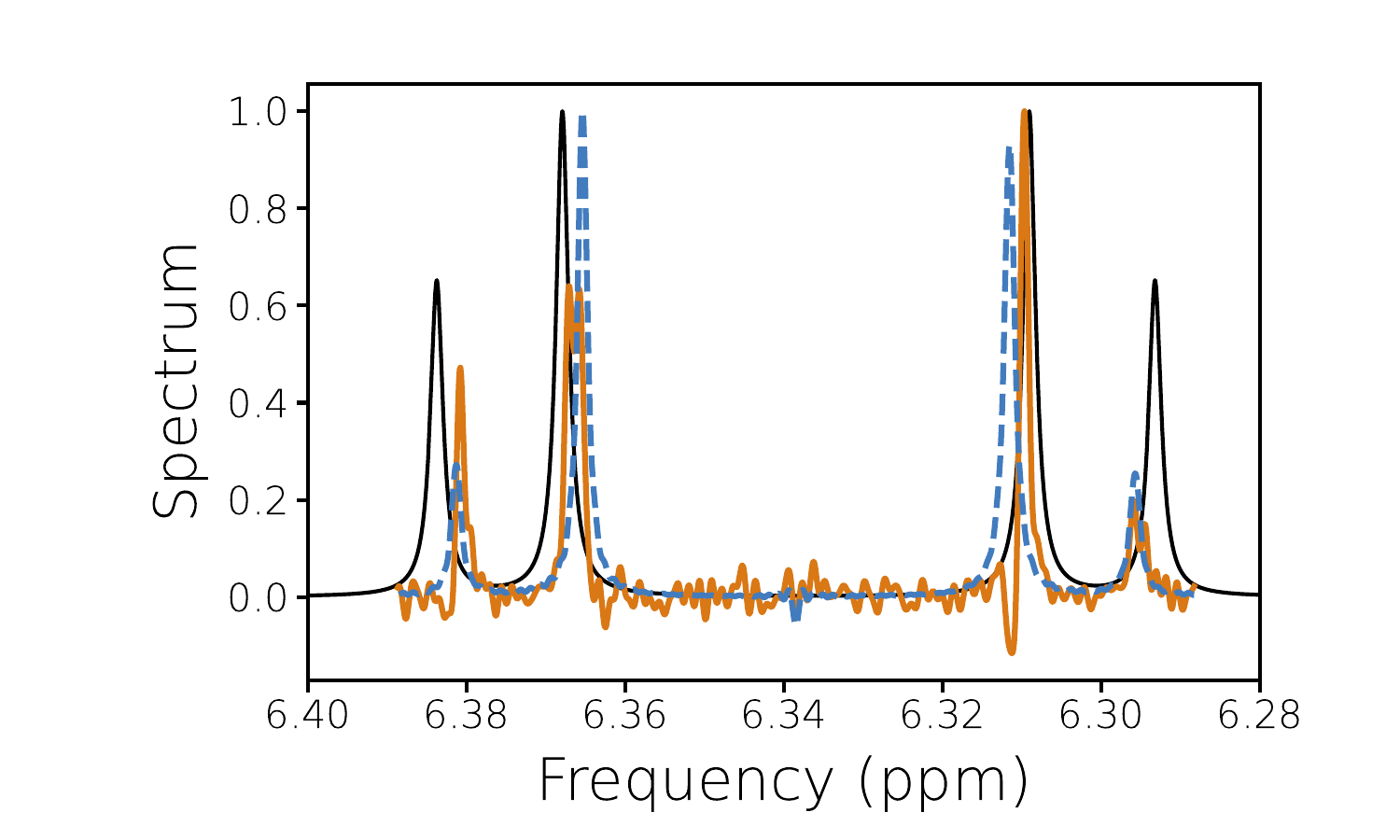}
    \caption{Simulation of NMR spectrum of \textit{cis}-3-chloroacrylic acid on IBM Perth with 81 Trotter steps of size $\tau=0.02$ shown by orange curve, compared with exact results (black curve) as well as simulation with on-gate depolarizig noise model (dashed blue curve).}
    \label{fig:spectrum_trotter_error}
\end{figure}

For a Trotterized time evolution, it is advantageous to increase the trotter step size to be able to simulate longer times,
and also to decrease the impact of noise, as discussed in the Method section.
However, by increasing the Trotter step size, we get an error from the Trotterization of the time evolution.
Fig.~\ref{fig:spectrum_trotter_error} depicts the occurrence of such an error in the form of shifts of the resonances.
Note that in the simulation with an on-gate depolarizing noise, we can reproduce such shifts to the actual resonances.

\section{Additional results}\label{app:additional_results}
\label{sec:appendixD}
In this section, we present further simulations of the NMR spectrum of \textit{cis}-3-chloroacrylic acid on IBM's processors. It is interesting to note that the deviations of the predicted resonances from the exact result varies in simulations performed on different days,
which can be a manifestation of quasistatic coherent noise.
\begin{figure}
    \centering
    \includegraphics[width=\columnwidth]{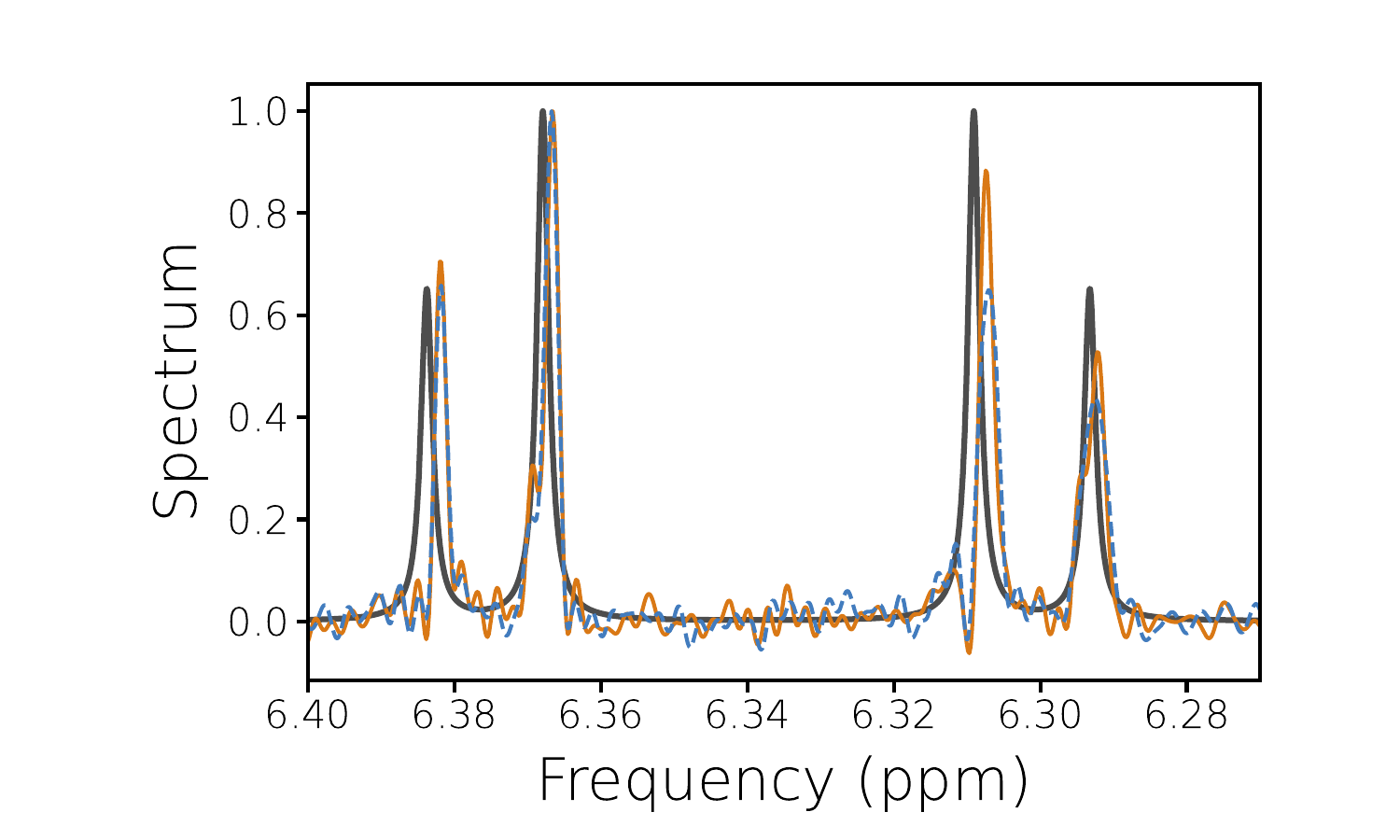}
    \caption{Comparison of the simulation of NMR spectrum of \textit{cis}-3-chloroacrylic acid on IBM Perth on three different days (shown with different colors) with exact results (black curve). For the simulation we performed a time evolution with 81 Trotter steps of size $\tau=0.01$.}
    \label{fig:spectrum_sym}
\end{figure}

\section{Effective decoherence rate}
\label{sec:appendixE}
\begin{figure}
    \centering
    \includegraphics[width=\columnwidth]{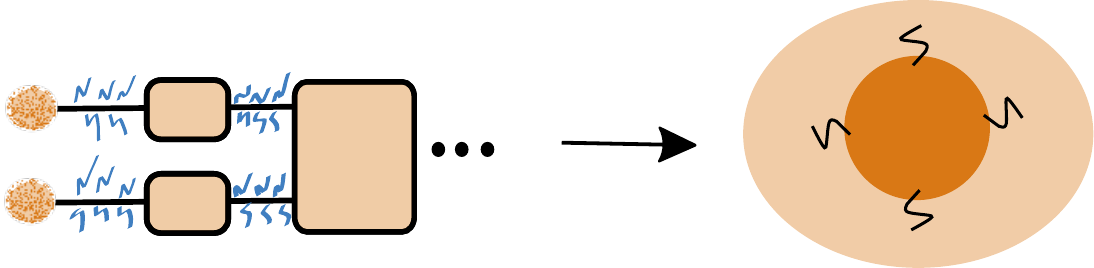}
    \caption{The sketch of noise mapping procedure outlined in Ref.~\cite{fratus2022describing}, where the noisy qubits and noisy erroneous gates involved in the implementation of the Trotterized time evolution result in an effective open quantum system described by Eq.~\ref{eq:App_lindblad}.}
    \label{fig:noise_mapping}
\end{figure}

In this section, we outline the details of extracting the effective decoherence rate from the equation of motion of spin-spin correlation function. As has been discussed in the main text such an effective decoherence rate allows us to find the precision with which the NMR spectrum is simulated on quantum chips. To this end, we first aim to find the open quantum system description of applying the Trotterized time evolution on a noisy quantum chip.
Such an open quantum system evolves according to master equation
\begin{align}
\dot{\rho} = \mathcal{L}\rho &= -i[H_\textrm{gates}(t),\rho]\nonumber \\
&+\sum_\alpha \gamma_{\alpha,\beta}\left(O_\alpha\rho O_\beta^\dagger -\frac{1}{2}\{O_\alpha O_\beta^\dagger, \rho\}\right),
\label{eq:App_lindblad}
\end{align}
where the matrix elements can be found from the noise mapping procedure explained in Ref.~\cite{fratus2022describing}, see Fig.~\ref{fig:noise_mapping}.
The spin operator like $\sigma^\alpha_i$, with $\alpha=x,y,z$ of the system then satisfies the following equation of motion
\begin{align}
\langle \dot{\sigma^\alpha_i} \rangle = &\text{Tr}\{O\dot{\rho}\}  = -i\text{Tr}\left\{[\sigma^\alpha_i,H_\textrm{gates}(t)]\rho\right\}\nonumber \\
&+\sum_\alpha \gamma_{\alpha}\text{Tr}\left\{\sigma^\alpha_i\left(O_\alpha\rho O_\alpha^\dagger -\frac{1}{2}\{O_\alpha O_\alpha^\dagger, \rho\}\right)\right\}.
\label{eq:evolution_sigma}
\end{align}
Working in the basis of Pauli operators for noise operators, we can show that
\begin{align}
\langle \dot{\sigma^\alpha_i} \rangle = \sum_{j} M_{ij} \langle \sigma^\alpha_i \rangle.
\end{align}
Moreover the quantum regression theorem
implies that
\begin{align}
\langle S^{z/y}_{\rm tot}(t) S^{z}_{\rm tot}\rangle&= \text{Tr}\left\{e^{iHt}S^{z/y}_{\rm tot}e^{-iHt}S^{z}_{\rm tot} \rho(0)\times \rho_{\rm B}\right\}\nonumber\\
&=\text{Tr}_{\rm s}\left\{S^{z/y}_{\rm tot} \text{Tr}_{\rm B}\left\{ e^{-iHt}S^{z}_{\rm tot} \rho(0)\times \rho_{\rm B} e^{iHt} \right\}\right\}\nonumber\\
&=\text{Tr}_{\rm s}\left\{S^{z/y}_{\rm tot} \mathcal{L} \left(S^{z}_{\rm tot}\rho(0)\right) \right\}
\end{align}
where we have assumed the density operator of the total system, i.e., system togeather with bath, is $\rho(0)\times \rho_{\rm B}$.
To better outline the implication of the regression theorem on the equation of motion of correlation functions, let us consider the case of one spin system $H=\epsilon\sigma_x$ with depopolarizing noise with rate $\Gamma_{\rm dep}$.
We can show that using cyclic properties of trace, Eq.\ref{eq:evolution_sigma} results in
\begin{align}
\langle \dot{\sigma^z} \rangle&=2\epsilon \langle \sigma^y \rangle -4 \Gamma_{\rm dep} \langle \sigma^z \rangle,\nonumber\\
\langle \dot{\sigma^y} \rangle&=-2\epsilon \langle \sigma^z \rangle -4 \Gamma_{\rm dep} \langle \sigma^y \rangle,\nonumber\\
\langle \dot{\sigma^x} \rangle&=2\epsilon \langle \sigma^x \rangle,
\end{align}
and for the correlation functions, it follows
\begin{align}
\frac{d}{dt}\langle \sigma^z(t) \sigma^z(0) \rangle &=-2i\epsilon \langle i\sigma_y(t)\sigma_z(0) \rangle-4\Gamma_{\rm dep} \langle \sigma_z(t)\sigma_z(0) \rangle \\
\frac{d}{dt}\langle i\sigma^y(t) \sigma^z(0) \rangle&=-2i\epsilon \langle i\sigma_z(t)\sigma_z(0) \rangle-4\Gamma_{\rm dep} \langle i\sigma_y(t)\sigma_z(0) \rangle
\end{align}
we can decouple the equations of motion for the linear combinations $\sigma^z\pm i \sigma^y$,
\begin{align}
&\frac{d}{dt}\langle [\sigma^z(t)\pm i\sigma^y(t)] \sigma^z(0) \rangle \nonumber\\
&= \left(\mp 2i\epsilon -4\Gamma_{\rm dep} \right) \langle[\sigma^z(t)\pm i\sigma^y(t)] \sigma^z(0)\rangle
\end{align}
which results in
\begin{align}
\langle\sigma^z(t)\sigma^z(0)\rangle=e^{-4\Gamma_{\rm dep}t}\bigg[&\cos(2\epsilon t) \langle \sigma^z(0)\sigma^z(0) \rangle \nonumber\\
&+ \sin(2\epsilon)\langle\sigma^y(t)\sigma^z(0)\rangle \bigg]
\label{eq:corr_1spin_ex}.
\end{align}
\begin{figure}
    \centering
    \includegraphics[width=\columnwidth]{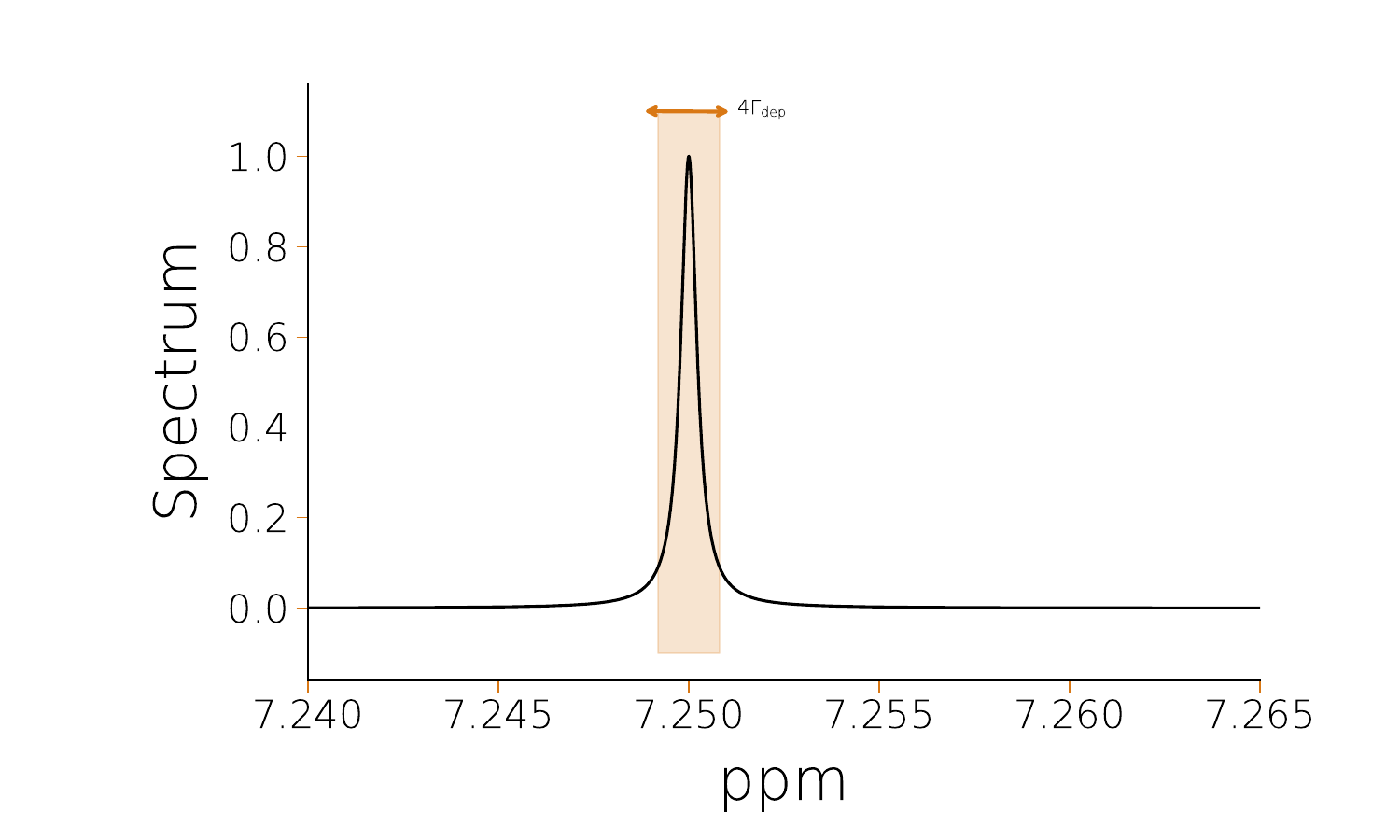}
    \caption{The NMR spectrum of $CHCl_3$ that has one spin for 1000 trotter steps of trotter time step $\tau = 0.005$. The shaded area shows the broadening $4\Gamma_{\rm dep}$ that indicates the decay of the correlation functions, see Eq.~\ref{eq:corr_1spin_ex}.}
    \label{fig:spectrum_sym}
\end{figure}
For larger systems, for liquid phase hamiltonian of Eq.~\ref{eq:Hamiltonian}, and depolarizing rate of strength $\Gamma_{\rm dep}$, we can show that
\begin{align}
\langle\dot{\sigma}^z_{\rm tot}\rangle=2i\sum_j \epsilon_j \langle\sigma^y_i\rangle -4 \Gamma_{\rm dep} \langle\sigma^z_{\rm tot}\rangle,
\label{eq:coupled_sigma_z}
\end{align}
\begin{align}
\langle\dot{\sigma}^y_{i}\rangle=&-2i\epsilon_i \langle\sigma^z_i\rangle -4 \Gamma_{\rm dep} \langle\sigma^y_{i}\rangle \nonumber\\
&+2i J\left[\langle \sigma^x_i\sigma^z_{j\neq i}\rangle-\langle \sigma^x_{j\neq i} \sigma_z^i\rangle \right],
\label{eq:coupled_sigma_y}
\end{align}
that indicates how the spin-spin coupling results in the coupling of equation of motion of spin operators with correlation functions. Ideally one has to solve the full set of coupled equations and find the decay of the correlation function which defines the effective decoherence rate. Note that while imaginary part  of the right-hand side of Eqs.~\ref{eq:coupled_sigma_z} and \ref{eq:coupled_sigma_y}, that results in oscillatory behaviour are coupled, the real part, resulting in the decay seems quite simple in the case of depolarizing noise. As in a generic case, finding the full solution of Eqs.~\ref{eq:coupled_sigma_z} and \ref{eq:coupled_sigma_y} is not feasible, we assume that such a coupling does not impact the decay rate, i.e., the effective decoherence rate. In the following we investigate to what extent such an assumption is valid by looking at the simulated spectrums of systems of various sizes.

\begin{figure}
    \centering
    \includegraphics[width=\columnwidth]{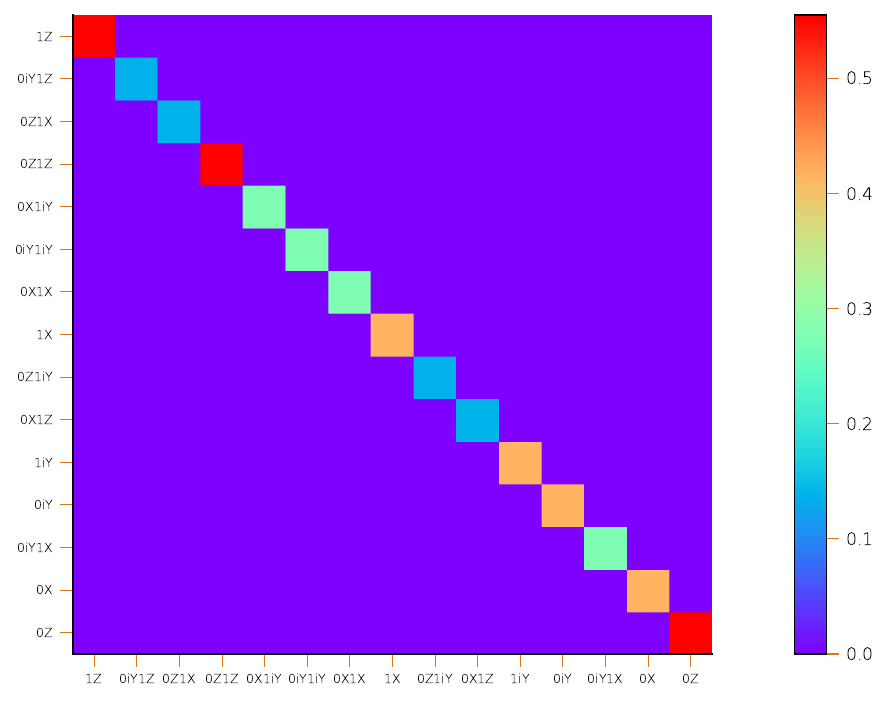}
    \caption{The matrix representation of decoherence rates $\gamma_{\alpha,\beta}$, see Eq.~\ref{eq:App_lindblad}, corresponding to a trotterized time evolution with $\tau=0.0025$ for cis-3-chloroacrylic acid.}
    \label{fig:noise_terms}
\end{figure}

Another complication that arises for larger systems, is that the effective open quantum system description is more complex than having a pure depolarizing noise. Note that even if we assume that the main processes describing the decoherence events for qubits are depolarizing noise, the presence of large angle gates, results in various complex noise terms. For instance, for the case of \textit{cis}-3-chloroacrylic acid, the open quantum system consists of single and two particle noise operators as shown in figure \ref{fig:noise_terms}.

For instance the noise term "iXiX" results in the following contribution
\begin{align}
\gamma_{iXiX}\left(\text{Tr}\{\sigma^z_k\sigma^x_i \sigma^x_i\}-\frac{1}{2}\text{Tr}\left\{\sigma^z_k \{\sigma^x_i\sigma^x_i,\rho\}\right\}\right)&\nonumber\\
=-2i\delta^{ik}\text{Tr}\{\sigma^y_i\sigma^x_i\}&
\label{eq:g_iXiX}
\end{align}
and similarly,
\begin{align}
\gamma_{iYiY}\left(\text{Tr}\left\{\sigma^z_k\sigma^y_i \sigma^y_i\right\}-\frac{1}{2}\text{Tr}\left\{\sigma^z_k \{\sigma^y_i\sigma^y_i,\rho\}\right\}\right)&\nonumber\\
=-2i\delta^{ik}\text{Tr}\{\sigma^x_i\sigma^y_i\}&
\label{eq:g_iYiY}
\end{align}
therefore if $\gamma_{iXiX}=\gamma_{iYiY}$, as is the case , see Fig.~\ref{fig:noise_terms}, we have Eq.~\ref{eq:g_iXiX}+Eq.~\ref{eq:g_iYiY}
$=-4\gamma_{iXiX}\delta^{ik}\langle \sigma^z_i\rangle $.

\begin{figure}
    \centering
    \includegraphics[width=\columnwidth]{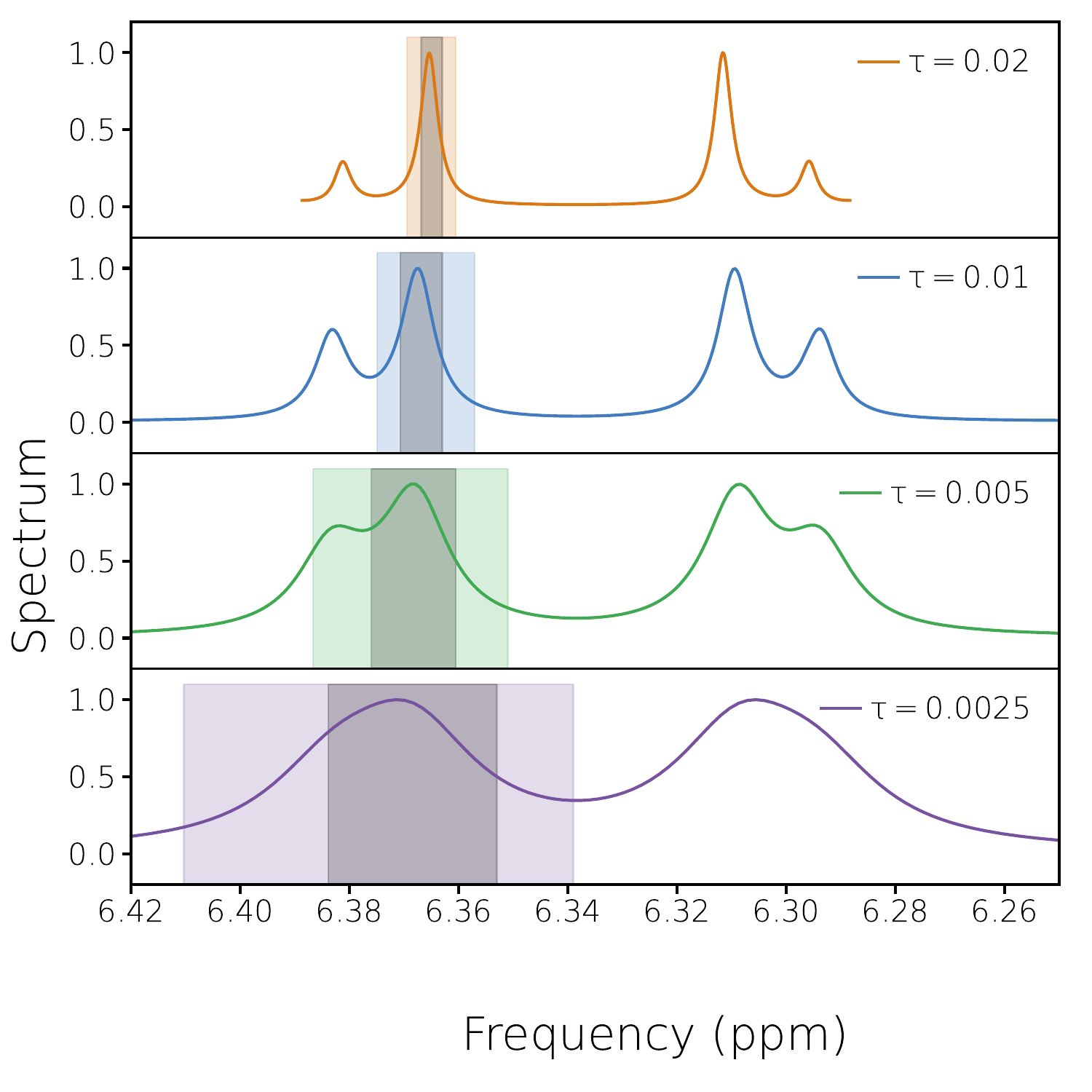}
    \caption{The simulated NMR spectrum of cis-3-chloroacrylic acid that has 2 spins for 1000 trotter steps of various trotter time
steps. The estimated broadening extracted from equation of motions is marked in different colors, while the definition Eq.~\ref{eq:effective_decoherence_rate}
is the gray area.}
    \label{fig:comparison_broadening_2spin}
\end{figure}

\begin{figure}
    \centering
    \includegraphics[width=\columnwidth]{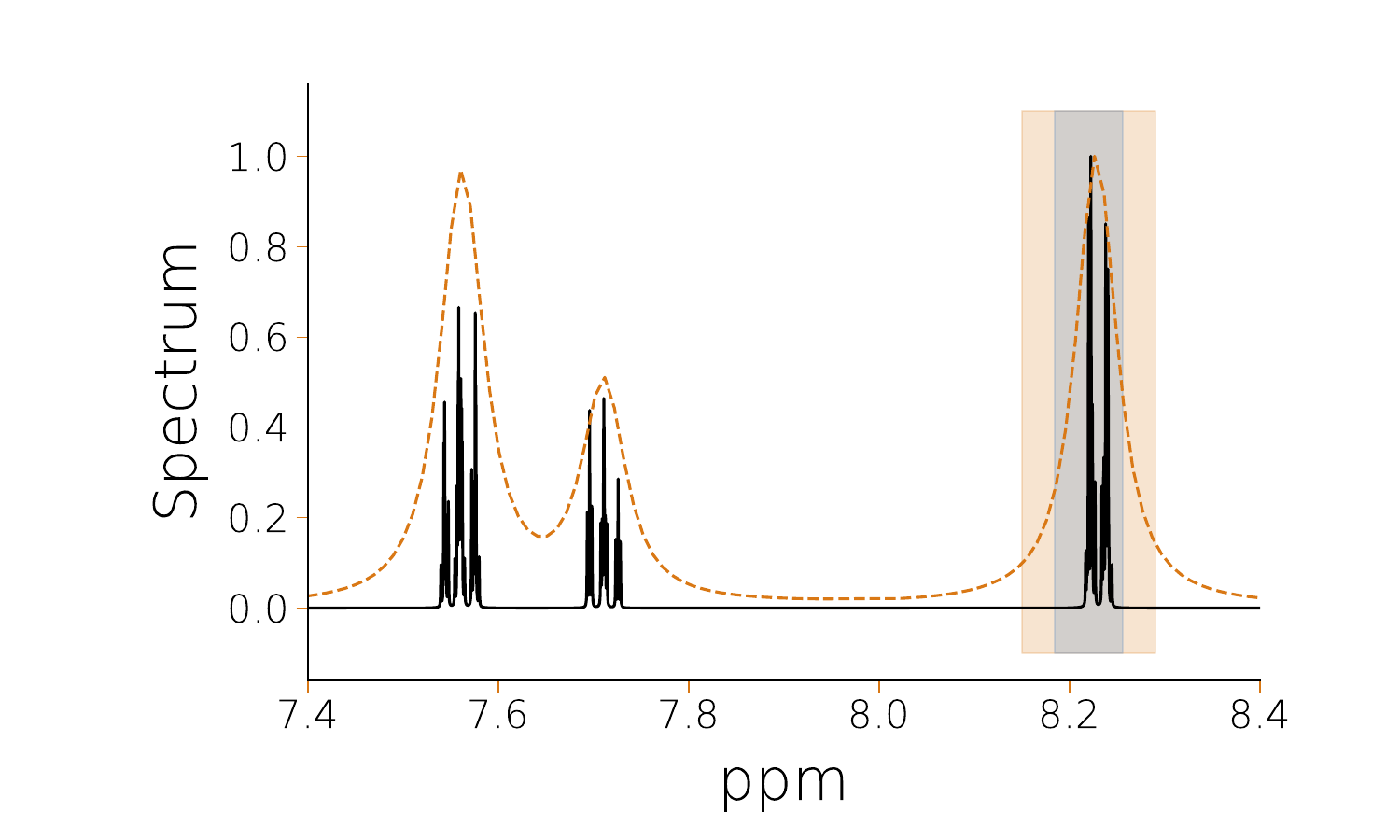}
    \caption{The simulated NMR spectrum of $C_6H_5NO_2$ that has 5 spins for 200 trotter steps of trotter time step $\tau = 0.001$. The
estimated broadening extracted from equation of motions is marked with orang, while the definition Eq.~\ref{eq:effective_decoherence_rate} is shown with light blue.}
   \label{fig:comparison_broadening_5spin}
\end{figure}

For completeness we compare the estimated broadening extracted from equation of motion with the definition from the main text Eq.~\ref{eq:effective_decoherence_rate}, where we consider a noisy simulation with on-gate depolarizing noise. As it is shown in Fig. \ref{fig:comparison_broadening_2spin} for a molecule with 2 spin, and in Fig.~\ref{fig:comparison_broadening_5spin} with a molecule that has 5 spins, the comparison indicates that the estimated broadening in the main text is a good measure of the precision of the digital simulation of NMR spectrum on NISQ devices.

\end{document}